\PassOptionsToPackage{dvipsnames,svgnames,x11names}{xcolor}
\documentclass{article} 
\usepackage[T1]{fontenc} 
\usepackage{iclr2027_conference,times}

\usepackage{amsmath,amsfonts,bm}

\def\eqref#1{equation~\ref{#1}}

\def\1{\bm{1}}

\DeclareMathAlphabet{\mathsfit}{\encodingdefault}{\sfdefault}{m}{sl}
\SetMathAlphabet{\mathsfit}{bold}{\encodingdefault}{\sfdefault}{bx}{n}

\usepackage{hyperref}
\usepackage{url}
\usepackage{graphicx}
\graphicspath{{../figures/}{figures/}}
\usepackage[dvipsnames,svgnames,x11names]{xcolor}
\usepackage{booktabs,colortbl,multirow,makecell,pifont,enumitem,wrapfig,listings}
\usepackage{subcaption} 
\definecolor{lprimary}{HTML}{714DC7}
\definecolor{lprimarylight}{HTML}{D4CAEC}
\definecolor{lprimarydark}{HTML}{4E328F}
\definecolor{lpurple}{HTML}{714DC7}
\definecolor{lpurplelight}{HTML}{D4CAEC}
\definecolor{lblue}{HTML}{4781C2}
\definecolor{lbluelight}{HTML}{CADAEC}
\definecolor{lteal}{HTML}{37A492}
\definecolor{lteallight}{HTML}{CAECE7}
\definecolor{lamber}{HTML}{C99136}
\definecolor{lamberlight}{HTML}{ECDFCA}
\definecolor{lpink}{HTML}{C55986}
\definecolor{lpinklight}{HTML}{ECCAD8}
\definecolor{lgray}{HTML}{8D99B0}
\definecolor{ltrunc}{HTML}{6B7788}
\definecolor{lgraylight}{HTML}{CAD7EC}
\definecolor{lheaderbg}{HTML}{F1EFF5}
\definecolor{lstripebg}{HTML}{ECECF0}
\definecolor{lborder}{HTML}{D7DAE0}
\definecolor{ltext}{HTML}{283548}
\definecolor{ltextsec}{HTML}{677389}
\definecolor{lbest}{HTML}{41AA7E}
\definecolor{lmid}{HTML}{C99136}
\definecolor{lworst}{HTML}{C35555}

\lstdefinestyle{prompt}{%
  basicstyle=\ttfamily\scriptsize\color{ltext},
  breaklines=true, breakautoindent=false, breakindent=0pt, columns=fullflexible,
  keepspaces=true, showstringspaces=false,
  aboveskip=0pt, belowskip=0pt,
  literate={±}{{$\pm$}}1 {–}{{--}}1 {秒}{{s}}1,
  moredelim=[il][\bfseries]{@@},
}

\hypersetup{
  colorlinks=true,
  citecolor=lblue,
  linkcolor=black,
  urlcolor=lblue,
}
\usepackage{etoc} 

\makeatletter
\renewcommand\paragraph{\@startsection{paragraph}{4}{\z@}{0.45ex plus 0.2ex minus .1ex}{-1em}{\normalsize\bf}}
\makeatother

\title{LongAudioSpan: Spanning the Duration and Depth of Audio Comprehension}

\author{%
Wen Huang$^{1\ast\dagger}$, Yunfei Chu$^{1\ast}$, Meng Gao$^{2\dagger}$, Haolin He$^{3\dagger}$, Jin Xu$^{1\ddagger}$\\
$^1$Qwen Team, Alibaba Group\quad $^2$Tsinghua University\quad$^3$The Chinese University of Hong Kong%
}

\newif\ifcameraready
\camerareadyfalse
\makeatletter
\newcommand{\cameraready}{%
  \iclrfinalcopy
  \camerareadytrue
  \gdef\@thanks{%
    \protect\footnotetext[1]{Equal contribution. $^{\dagger}$Work done during internship at Alibaba Group. $^{\ddagger}$Corresponding author.}}%
}
\makeatother
\cameraready   

\newcommand{\coderepo}{\ifcameraready https://huggingface.co/datasets/holvan/LongAudioSpan\else https://huggingface.co/datasets/audiospan/LongAudioSpan\fi}

\newcommand{\benchmark}{LongAudioSpan}

\newcommand{\yes}{\textcolor{lteal}{\ding{51}}}
\newcommand{\no}{\textcolor{lpink}{\ding{55}}}
\newcommand{\half}{\textcolor{lamber}{\ding{119}}}

\newcommand{\best}[1]{\textbf{#1}}
\newcommand{\second}[1]{\underline{#1}}
\newcommand{\rowstripe}{\rowcolor{lstripebg}}
\newcommand{\bl}[1]{\textcolor{ltrunc}{#1}}
\newcolumntype{A}{>{\cellcolor{lheaderbg}}wc{0.6cm}}
\newsavebox{\fitbox}
\newcommand{\fittable}[1]{%
  \sbox{\fitbox}{#1}%
  \ifdim\wd\fitbox>\textwidth
    \resizebox{\textwidth}{!}{\usebox{\fitbox}}%
  \else
    \usebox{\fitbox}%
  \fi}

\usepackage{alltt}
\usepackage{tikz}
\usetikzlibrary{shadows}
\usepackage{tcolorbox}
\tcbuselibrary{skins,breakable,listings}
\colorlet{cbaccent}{lprimary}
\newcommand{\cbhd}[1]{\textbf{#1}}
\newsavebox{\casebodybox}\newsavebox{\cbtitlebox}
\newlength{\cbwd}\newlength{\cbpad}\newlength{\cbtpad}\newlength{\cbrad}
\newlength{\cbW}\newlength{\cbHdrH}\newlength{\cbBodH}\newlength{\cbTotH}
\newenvironment{casebox}[2][blue]{%
  \colorlet{cbaccent}{l#1}%
  \setlength{\cbwd}{\dimexpr\linewidth-2\cbpad\relax}
  \sbox{\cbtitlebox}{\bfseries\itshape\footnotesize\textcolor{white}{#2}}%
  \par\addvspace{\medskipamount}%
  \begin{lrbox}{\casebodybox}%
  \begin{minipage}{\cbwd}%
    \ttfamily\scriptsize\color{ltext}%
}{%
  \end{minipage}%
  \end{lrbox}%
  \setlength{\cbW}{\dimexpr\cbwd+2\cbpad\relax}%
  \setlength{\cbHdrH}{\dimexpr\ht\cbtitlebox+\dp\cbtitlebox+2\cbtpad\relax}%
  \setlength{\cbBodH}{\dimexpr\ht\casebodybox+\dp\casebodybox+2\cbpad\relax}%
  \setlength{\cbTotH}{\dimexpr\cbHdrH+\cbBodH\relax}%
  \par\noindent\begin{center}%
  \begin{tikzpicture}%
    \node[anchor=south west, inner sep=0pt, minimum width=\cbW, minimum height=\cbTotH,
          rounded corners=\cbrad, fill=white, draw=cbaccent, line width=0.6pt,
          drop shadow={shadow xshift=1pt, shadow yshift=-1pt, opacity=0.16, fill=lgray}]
         at (0,0) {};%
    \begin{scope}%
      \clip[rounded corners=\cbrad] (0,0) rectangle (\cbW,\cbTotH);%
      \fill[cbaccent] (0,\cbBodH) rectangle (\cbW,\cbTotH);%
    \end{scope}%
    \draw[rounded corners=\cbrad, draw=cbaccent, line width=0.6pt] (0,0) rectangle (\cbW,\cbTotH);%
    \node[anchor=west, inner sep=0pt] at (\cbpad,\dimexpr\cbTotH-\cbHdrH/2\relax) {\usebox{\cbtitlebox}};%
    \node[anchor=north west, inner sep=0pt] at (\cbpad,\dimexpr\cbBodH-\cbpad\relax) {\usebox{\casebodybox}};%
  \end{tikzpicture}%
  \end{center}\addvspace{\smallskipamount}}

\newtcblisting{promptbox}[1]{%
  breakable, enhanced,
  listing only, listing options={style=prompt},
  colback=white, colframe=lteal, boxrule=0.8pt, arc=\cbrad,
  colbacktitle=lteal, coltitle=white, fonttitle=\bfseries\itshape\footnotesize,
  boxsep=0pt, titlerule=0pt,
  title={#1},
  left=\cbpad, right=\cbpad, top=\cbpad, bottom=\cbpad,
  toptitle=\cbtpad, bottomtitle=\cbtpad,
}

\begin{document}

\maketitle
\ificlrfinal
\lhead{Under review}
\fi

\etocsettocdepth.toc{none}

\begin{abstract}
General audio comprehension now covers speech, sound, and music over durations from seconds to hours, driven by large audio-language models (LALMs) that are increasingly omni-modal. Yet the benchmarks that test them still rely on clips of seconds, where scores saturate and models converge; recent long-form efforts extend duration but evaluate long audio much as short clips are. We introduce \textbf{\benchmark}, a benchmark that spans both duration and depth: it pairs audio from 10 minutes to over 2 hours with 3{,}240 questions across three cognitive levels, namely perception, understanding, and reasoning. Two paths supply the questions, differing in how question content is sourced and how ground truth is obtained. Native QA extracts questions from the audio's content, posing each as a multiple-choice item and an open-ended one graded by detailed rubrics. Anchor QA instead injects ground truth, planting acoustic anchors into the audio and building a perception-to-reasoning chain scored only to the first error. A fully automated pipeline constructs every item through structured captioning, QA generation, and adversarial critic feedback. Evaluating 12 LALMs on \benchmark, we find the hard part comes before reasoning: distilling a few relevant facts from a long, redundant signal. This difficulty grows with audio length and falls hardest on perception, especially temporal grounding. \benchmark\ is available at \url{\coderepo}.
\end{abstract}

\section{Introduction}
\label{sec:intro}

Audio intelligence has expanded from specialized recognition tasks to encompass general audio-language comprehension. Early work focused on independent tasks such as speech recognition \citep{radford2023whisper}, sound classification \citep{chen2023beats}, and audio-text alignment \citep{wu2022large}, each relying on specialized architectures with fixed output formats. More recently, large audio-language models (LALMs), from dedicated audio specialists to omni-modal systems that also handle vision and text, unify diverse audio capabilities through open-ended text generation. Over the past three years, these models have evolved along several axes: from speech-only \citep{zhang2023speechgpt} to unified audio covering speech, sound, and music \citep{chu2023qwen}, from simple recognition to open-ended understanding \citep{chu2024qwen}, and from short clips \citep{deshmukh2023pengi,tang2023salmonn} to multi-hour audio \citep{qwen2026omni}.

Benchmarks for evaluating LALMs have emerged \citep{sakshi2025mmau,ma2025mmar,wang2026mmsu}, covering multidimensional assessment across different audio modalities, with task types from basic perception to deeper reasoning. As model capabilities improve, top accuracy on these benchmarks has risen from around 50\% to over 80\% within a year of publication, and the leading LALMs now cluster within a few points of one another. These benchmarks, however, operate on audio clips measured in seconds: the relevant evidence sits in a short span, and answering does not call for sustained attention across a long context.

\begin{table}[t!]
\centering
\caption{Comparison of \benchmark\ with existing audio comprehension benchmarks.}
\label{tab:benchmark-comparison}
\setlength{\tabcolsep}{5pt}
\renewcommand{\arraystretch}{1.1}
\fittable{%
\footnotesize
\begin{tabular}{lrwc{1.1cm}wc{1.1cm}wc{1.1cm}wc{1.1cm}wc{1.1cm}wc{1.1cm}wc{1.1cm}wc{1.1cm}}
\toprule
  & & \multicolumn{2}{c}{\textbf{Task Framework}}
  & \multicolumn{2}{c}{\textbf{Question Format}}
  & \multicolumn{2}{c}{\textbf{Data Construction}}
  & \multicolumn{2}{c}{\textbf{Quality Assurance}} \\[2pt]
\cmidrule(lr){3-4} \cmidrule(lr){5-6} \cmidrule(lr){7-8} \cmidrule(lr){9-10}
\textbf{Benchmark}
  & \textbf{Duration} & \textbf{Depth} & \textbf{Scheme}
  & \textbf{Choice} & \textbf{Open}
  & \textbf{Source} & \textbf{Pipeline}
  & \textbf{Filter} & \textbf{Refine} \\
\midrule
MMSU \citep{wang2026mmsu}
  & {$\sim$7\,s} & Tiered & Fixed
  & \yes & \no
  & Wild & Manual
  & Manual & None \\
\rowstripe MMAU \citep{sakshi2025mmau}
  & {$\sim$10\,s} & Tiered & Fixed
  & \yes & \no
  & Corpus & Manual
  & Manual & None \\
MMAR \citep{ma2025mmar}
  & {$\sim$20\,s} & Tiered & Fixed
  & \yes & \no
  & Wild & Manual
  & Manual & None \\
\rowstripe MMAU-Pro \citep{kumar2025mmau}
  & {$\sim$2\,min} & Tiered & Fixed
  & \yes & \yes
  & Wild & Manual
  & Manual & None \\
LongAudioBench \citep{ghosh2025audio}
  & {$\sim$2\,min} & Flat & Fixed
  & \yes & \yes
  & Corpus & Auto
  & Multi & None \\
\rowstripe AudioMarathon \citep{he2025audiomarathon}
  & {$\sim$3\,min} & Flat & Fixed
  & \yes & \half
  & Corpus & Auto
  & Manual & None \\
ChronosAudio \citep{luo2026chronosaudio}
  & {$\sim$6\,min} & Tiered & Fixed
  & \no & \half
  & Synthetic & Auto
  & None & None \\
\rowstripe LongSpeech \citep{yang2026longspeech}
  & {$\sim$10\,min} & Flat & Fixed
  & \no & \half
  & Corpus & Auto
  & Manual & None \\
LAT-Bench \citep{shao2026latbench}
  & {$\leq$30\,min} & Flat & Fixed
  & \no & \half
  & Wild & Auto
  & None & None \\
\rowstripe BLAB \citep{ahia2025blab}
  & {$\sim$51\,min} & Flat & Fixed
  & \yes & \half
  & Wild & Manual
  & Manual & None \\
VoiceGiraffe \citep{ye2026voicegiraffe}
  & {$\sim$55\,min} & Tiered & Fixed
  & \yes & \no
  & Wild & Auto
  & Multi & None \\
\midrule
\textbf{\benchmark\ (Ours)}
  & \makebox[0pt][r]{{\textbf{10\,min--2\,h$^+$ \,$\|$\, $\sim$50\,min}}}
  & \textbf{Tiered} & \textbf{Free}
  & \textbf{\yes} & \textbf{\yes}
  & \textbf{Wild} & \textbf{Auto}
  & \textbf{Multi} & \textbf{Adaptive} \\
\bottomrule
\multicolumn{10}{@{}l@{}}{\scriptsize\textcolor{ltextsec}{%
In \textbf{Open}: \yes\ open-ended question;\ \half\ fixed-answer question (e.g., automatic speech recognition, ASR).}} \\
\end{tabular}%
}
\end{table}

Recent efforts have begun to fill this gap by extending audio duration to the minute level \citep{yang2026longspeech,luo2026chronosaudio,ghosh2025audio,he2025audiomarathon,shao2026latbench} or beyond \citep{ahia2025blab,ye2026voicegiraffe} (Table~\ref{tab:benchmark-comparison}). Yet even with extended duration, these benchmarks share limitations across four dimensions: \textbf{1)~Task framework}: tasks follow fixed schemes over a predetermined set of types, and cognitive depth is mostly flat. \textbf{2)~Question format}: most rely on multiple-choice or fixed-answer questions, where models can achieve inflated scores through shortcut techniques rather than genuine comprehension. \textbf{3)~Data construction}: synthetic or corpus-derived audio yields reliable ground truth but lacks real-world variation; wild audio either requires costly manual annotation or depends on caption quality through caption-then-QA pipelines. \textbf{4)~Quality assurance}: existing pipelines typically rely on post-hoc human review as a single filtering layer, with no mechanism to adaptively refine questions as the benchmark evolves.

To address these gaps, we introduce \emph{\textbf{\benchmark}}, a benchmark spanning both duration, from minutes to hours, and depth of audio comprehension through a three-level cognitive framework of \emph{perception}, \emph{understanding}, and \emph{reasoning}. To construct \benchmark, we collect long-form audio in the wild and design a fully automated, scalable pipeline in two phases: Phase~1 generates a structured caption from audio; Phase~2 generates QA pairs with multi-level quality control and adaptive refinement. It follows two complementary paths: \textbf{Native QA}, which derives questions natively from the audio's inherent content, freely composing diverse question types within the cognitive framework, in both multiple-choice and open-ended formats; and \textbf{Anchor QA}, which introduces acoustic anchors into the audio and builds a question chain centered on the anchor and its surrounding context, with chain scoring to suppress short-cuts.

We evaluate \benchmark\ on 12 recent LALMs, seven open-source and five proprietary. Even the strongest models are bottlenecked by distilling a few relevant facts from a long, redundant signal before any reasoning begins, a difficulty that grows with audio length.

In summary, we make the following contributions:

\begin{itemize}[leftmargin=1.5em, itemsep=1.5pt, parsep=0.45ex plus 0.2ex minus 0.1ex]
\item \textbf{A comprehensive long-form audio benchmark with diverse evaluation paradigms} (\S\ref{sec:design}). \benchmark\ pairs bilingual in-the-wild audio spanning minutes to hours with a dual-path evaluation across three cognitive levels and three question formats.

\item \textbf{A scalable automated construction pipeline with multi-level quality assurance} (\S\ref{sec:construction}). We automate the full loop from question generation to filtering, intercepting low-quality items at generation time and feeding adversarial critic feedback into later rounds.

\item \textbf{A systematic empirical study of long-form audio comprehension} (\S\ref{sec:experiments}). Across a broad set of recent LALMs, we analyze long-form comprehension along duration, cognitive levels, and scoring modes, characterizing where and why it breaks down.
\end{itemize}

\section{Related Work}
\label{sec:related}

\subsection{Large Audio-Language Models}
\label{sec:related-lalms}

Audio-language modeling has progressed from modality-specific systems toward unified, general-purpose models. Early systems targeted individual modalities and tasks: speech-centric language models \citep{zhang2023speechgpt,rubenstein2023audiopalm} and general audio encoders coupled to LLMs for captioning and question answering over short clips \citep{deshmukh2023pengi,tang2023salmonn,gong2023listen,ghosh2024gama}. The Qwen-Audio line \citep{chu2023qwen,chu2024qwen} brought speech, environmental sound, and music into a single model, and open releases have since proliferated along two lines. Audio specialists keep the input interface audio-only: the Audio Flamingo series \citep{goel2025audio,ghosh2026audio}, Voxtral \citep{mistral2025voxtral}, Kimi-Audio \citep{kimiteam2025kimi}, MiMo-Audio \citep{team2025mimo}, StepAudio \citep{lin2026stepaudio}, Eureka-Audio \citep{zhang2026eureka}, and MOSS-Audio \citep{openmoss2026moss}. Omni-modal models pair audio with vision and text: the Qwen-Omni series \citep{xu2025qwen,xu2025qwen1}, Phi-4-Multimodal \citep{abouelenin2025phi}, Baichuan-Omni \citep{li2025baichuan}, and MiniCPM-o \citep{cui2026minicpm}. Recent releases also push toward longer effective context \citep{xu2025qwen1,qwen2026omni}, stronger multi-step reasoning \citep{xie2025audio,tian2025step}, and finer temporal grounding \citep{sun2026spotsound,hegde2026event}. Alongside open models, proprietary systems such as the Gemini series \citep{team2024gemini,geminiteam2025gemini,geminiteam2026gemini31} and Qwen3.5-Omni \citep{qwen2026omni} provide native audio understanding over long contexts.

\subsection{Audio Comprehension Benchmarks}
\label{sec:related-benchmarks}

Benchmarks for LALMs initially centered on short clips under one minute. MMAU \citep{sakshi2025mmau}, AudioBench \citep{wang2025audiobench}, and AIR-Bench \citep{yang2024air} assess perception and understanding across speech, sound, and music, while MMAU-Pro \citep{kumar2025mmau} broadens task coverage and adds open-ended items. MMAR \citep{ma2025mmar}, MMSU \citep{wang2026mmsu}, and STAR-Bench \citep{liu2026starbench} instead target deeper reasoning, multi-step or spatio-temporal, but still on clips measured in seconds. A more recent line pushes audio from a few minutes to about an hour (Table~\ref{tab:benchmark-comparison}), drawing on corpora or synthetic narration (LongAudioBench \citep{ghosh2025audio}, AudioMarathon \citep{he2025audiomarathon}, ChronosAudio \citep{luo2026chronosaudio}, LongSpeech \citep{yang2026longspeech}) or in-the-wild recordings (LAT-Bench \citep{shao2026latbench}, BLAB \citep{ahia2025blab}, VoiceGiraffe \citep{ye2026voicegiraffe}). Yet evaluation still treats long audio much as it does short clips, with fixed task schemes and mostly flat depth, and construction remains largely automated over synthetic or caption-derived ground truth and checked by one-off filters.

\subsection{QA Generation Pipelines}
\label{sec:related-qa}

Constructing audio QA at scale requires generating questions whose answers are grounded in the audio. Early benchmarks rely on expert annotation \citep{sakshi2025mmau,ma2025mmar,wang2026mmsu}, which yields reliable ground truth but scales poorly. To reduce annotation cost, many recent pipelines adopt a caption-then-QA design, deriving questions from a textual description of the audio rather than from the audio itself \citep{ye2026voicegiraffe,ghosh2025audio,wang2025audiobench}. This design scales readily, but the resulting questions are bounded by the fidelity of the intermediate caption. Pipelines also differ in how they guard quality, from full human verification \citep{sakshi2025mmau} to LLM-based filtering \citep{ye2026voicegiraffe} and post-hoc text-only audits that flag items answerable without the audio \citep{kumar2025mmau}. These checks are one-off filters applied after generation.

\subsection{Evaluation Methods}
\label{sec:related-eval}

Audio benchmarks predominantly score with multiple-choice questions, which afford deterministic grading (a 25\% random baseline for four options) but can be solved by option elimination rather than genuine comprehension. Open-ended items fall into two kinds: fixed-answer questions keep grading automatic but constrain what can be asked, while free-form questions better reflect real usage and rely on LLM-as-judge scoring \citep{kumar2025mmau,ghosh2025audio,luo2026chronosaudio}. Such judging is typically holistic, a single overall rating that does not show which aspects of a response are correct; finer criteria have only graded reasoning traces behind multiple-choice answers \citep{ma2025mmar}. Chain-structured evaluation, explored in video understanding, links successive questions into a dependent sequence \citep{fu2026video}.

\section{Benchmark Design}
\label{sec:design}

\subsection{Overview}
\label{sec:design-overview}

\benchmark's design centers on two complementary question paths that differ in how question content is sourced and how ground truth is obtained. The base path, \textbf{Native QA}, draws questions from the audio's original content via a structured caption of the audio (\S\ref{sec:construction}). Its ground truth, however, is bounded by caption completeness: content omitted from the caption cannot be turned into questions, and on hour-long audio such omissions are hard to avoid. To sidestep this dependency, the second path, \textbf{Anchor QA}, injects ground truth rather than extracting it: it plants acoustic anchors into the audio and builds questions around their surrounding context, so the operation itself defines the correct answer. Together, the two paths combine broad coverage of natural content with verifiable ground truth at targeted positions.

The two paths share a unified cognitive taxonomy but differ in question format and scoring (\S\ref{sec:design-taxonomy}). Instantiated over 360 in-the-wild audio samples, they yield 3,240 QA items balanced across languages (English and Chinese) and duration tiers (10 min to 2 h$^+$) (\S\ref{sec:design-stats}).

\subsection{Taxonomy and Evaluation}
\label{sec:design-taxonomy}

\begin{figure}[t]
\centering
\includegraphics[width=\textwidth]{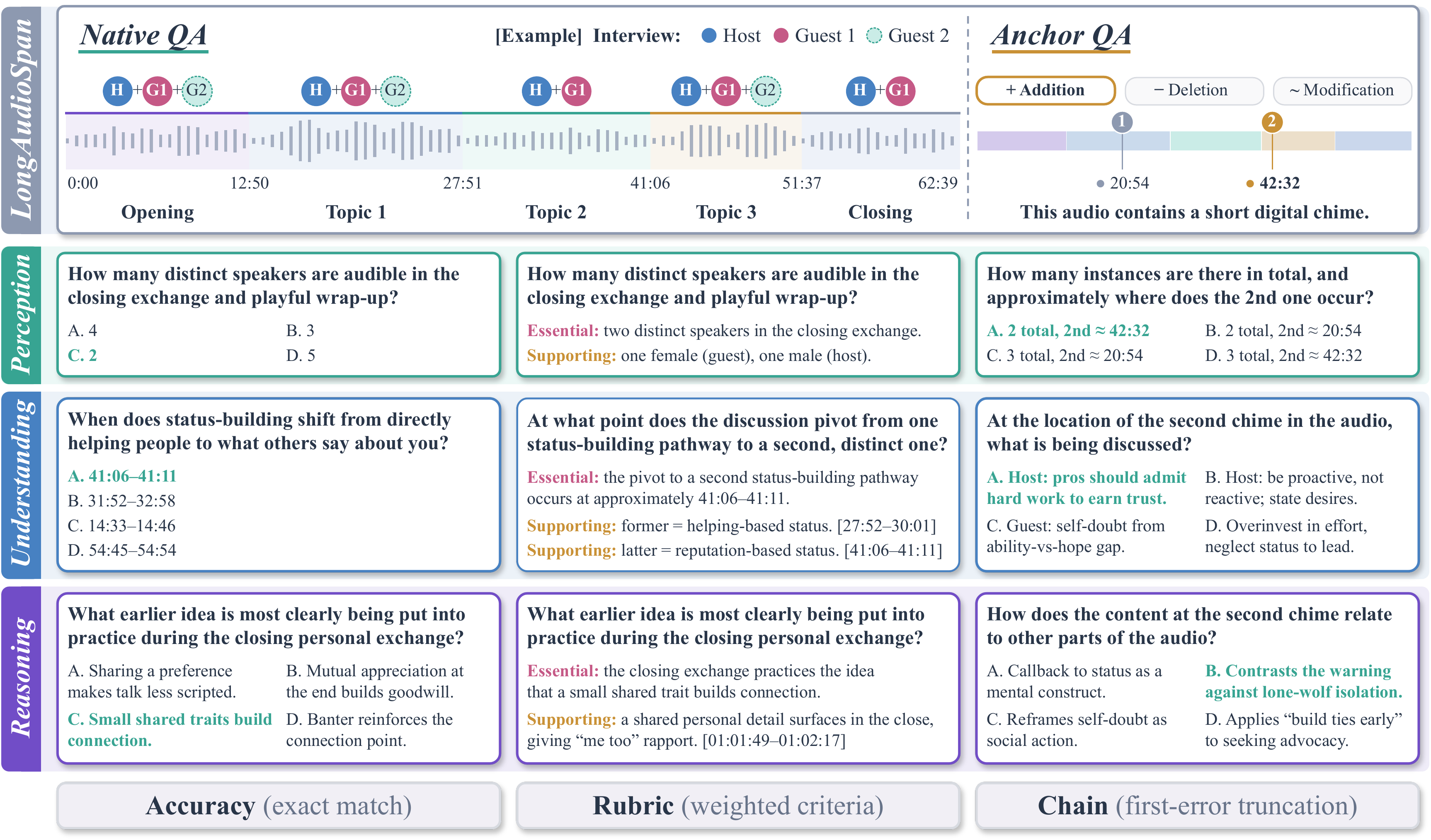}
\caption{Overview of \benchmark. Top: the two question paths over one audio sample, Native QA (left) and Anchor QA (right). Bottom:  abridged example question for each taxonomy level (rows) and scoring scheme (columns).}
\label{fig:design-overview}
\end{figure}

Figure~\ref{fig:design-overview} presents each path's questions and scoring built on a shared taxonomy. The taxonomy follows a cognitive progression from information acquisition to inferential judgment. At the base, \emph{Perception} (P) involves identifying and acquiring explicit information present in the audio. \emph{Understanding} (U) constructs meaning from that explicit information, building coherent interpretations of the content. \emph{Reasoning} (R) draws inferences and judgments that go beyond the given information.

\paragraph{Native QA.} Native QA realizes the taxonomy as a free scheme. Each level is organized into three general dimensions, each opening a question space (Appendix~\ref{app:taxonomy}). For each audio, the generator draws on the question space for inspiration and grounds each question in the audio's content, yielding questions specific to that audio. Each question is posed in both formats, as an MCQ and an OEQ. Multiple-choice questions (MCQs) with four options are scored by exact match, the \emph{Accuracy} over an MCQ set $Q$:
\begin{equation}
\label{eq:accuracy}
\mathrm{Accuracy} = \frac{1}{|Q|} \sum_{q \in Q} \mathbf{1}[\hat{a}_q = a_q].
\end{equation}
Open-ended questions (OEQs) are graded by an ensemble of rubric-based LLM judges against item-specific criteria. Each criterion $c$ takes a \emph{semantic} score $\sigma_c \in \{0, 0.5, 1\}$ for content correctness and, when the criterion is time-bound, a \emph{temporal} score $\tau_c \in \{0, 1\}$ that grounds the content at the right moment, treating a timestamp as a citation. The \emph{Rubric} score is the importance-weighted mean of the per-criterion scores $s_c$:
\begin{equation}
\label{eq:rubric}
\mathrm{Rubric} = \frac{\sum_{c \in C} w_c\, s_c}{\sum_{c \in C} w_c},
\qquad
s_c =
\begin{cases}
\sigma_c \text{ or } \tau_c, & \text{essential,}\\[2pt]
\mathbf{1}[\sigma_c > 0]\,(\sigma_c + \tau_c)/2, & \text{supporting.}
\end{cases}
\end{equation}
An \emph{essential} criterion checks the direct answer to the question and is scored on a single dimension, $\sigma_c$ for semantic content or $\tau_c$ for temporal localization, at weight $w_c = 1$. A \emph{supporting} criterion rewards a detail that substantiates the answer at weight $w_c = 0.5$, scored on semantic content with temporal grounding as an additional trait. Grounding is conditional on content: a timestamp acts as a citation, so it earns no credit for a false statement, and the indicator $\mathbf{1}[\sigma_c > 0]$ zeros a supporting criterion whose content is wrong (Appendix~\ref{app:rubric}). The OEQ withholds the options the MCQ provides, removing the elimination shortcut and measuring a model's ability more realistically.

\paragraph{Anchor QA.} Anchor QA uses a fixed chain that maps directly onto P$\to$U$\to$R. Each audio sample undergoes one of three anchor operations: \emph{Addition} inserts a sound event into the audio, \emph{Deletion} applies a silence or noise mask to remove information at chosen positions, and \emph{Modification} alters acoustic attributes such as volume or playback speed (Appendix~\ref{app:anchor}). Around each anchor we construct three MCQs, one per level: P grounds the anchor by counting its occurrences and locating the Nth instance; U then infers the content being discussed there; and R in turn relates that content to other parts of the audio. The three questions form an implicit cognitive dependency, so an answer that follows an earlier error rests on a wrong premise. The \emph{Chain} score therefore credits answers only up to the first error,
\begin{equation}
\label{eq:chain}
\mathrm{Chain} = \frac{1}{3} \sum_{k=1}^{3} \prod_{j=1}^{k} \mathbf{1}[\hat{a}_j = a_j],
\end{equation}
where $\hat{a}_j$ is the answer to the $j$-th question, so one early mistake zeros every later term. Under random guessing the expected score is 10.9\%, well below the 25\% baseline of a single MCQ.

\begin{figure}[t]
\centering
\noindent\hspace*{-10pt}\makebox[\textwidth]{%
\begin{minipage}[t]{0.46\textwidth}
  \centering
  \includegraphics[width=\textwidth]{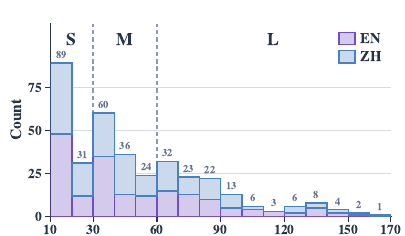}\\[2pt]
  {\small (a) Duration (min)}
\end{minipage}\hfill
\begin{minipage}[t]{0.245\textwidth}
  \centering
  \includegraphics[width=\textwidth]{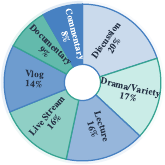}\\[2pt]
  {\small (b) Genre}
\end{minipage}\hfill
\begin{minipage}[t]{0.245\textwidth}
  \centering
  \includegraphics[width=\textwidth]{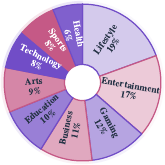}\\[2pt]
  {\small (c) Topic}
\end{minipage}%
}
\caption{Distribution of source audio across (a) duration tiers, (b) genres, and (c) topics.}
\label{fig:distribution}
\end{figure}

\subsection{Statistics}
\label{sec:design-stats}

\begin{wraptable}{r}[8pt]{4.2cm}
\vspace{-1.9\baselineskip}
\caption{Benchmark statistics.}
\label{tab:stats}
\fontsize{8}{9.5}\selectfont
\setlength{\tabcolsep}{4pt}
\begin{tabular}{lr}
\toprule
\textbf{Statistic} & \textbf{Value} \\
\midrule
\rowcolor{lheaderbg} \textbf{\textit{Audio Sources}} & 720 \\[2pt]
\quad Original & 360 \\
\qquad Per tier (S/M/L) & 120 \\
\qquad Per lang. (EN/ZH) & 180 \\
\quad With anchors & 360 \\
\qquad Per mode (A/D/M) & 120 \\
\addlinespace
\rowcolor{lheaderbg} \textbf{\textit{QA Items}} & 3,240 \\[2pt]
\quad Per audio (1P/1U/1R) & 3 \\
\quad Native (MCQ) & 1,080 \\
\quad Native (OEQ) & 1,080 \\
\quad Anchor (MCQ) & 1,080 \\
\bottomrule
\end{tabular}
\end{wraptable}

\benchmark\ draws on 360 in-the-wild audio samples, split evenly across three duration tiers, S (10--30 min), M (30--60 min), and L (60+ min), and between English and Chinese; the samples further span 7 genres and 9 topics (Figure~\ref{fig:distribution}). Every sample also has an anchor-modified version under one of three operations, for 720 audio files totaling roughly 600 hours (avg. $\sim$50\,min). The tiers form a deliberate duration gradient: S, M, and L average roughly 16, 42, and 90 minutes, with the longest sample reaching 2.7 hours.

Each sample is annotated with one question per cognitive level under each evaluation format, yielding 3,240 items: 1,080 Native MCQs, 1,080 Native OEQs, and 1,080 Anchor MCQs organized as 360 three-question chains (Table~\ref{tab:stats}).

\section{Data Construction}
\label{sec:construction}

\subsection{Overview}
\label{sec:construction-overview}

\begin{figure}[t]
\centering
\includegraphics[width=\textwidth]{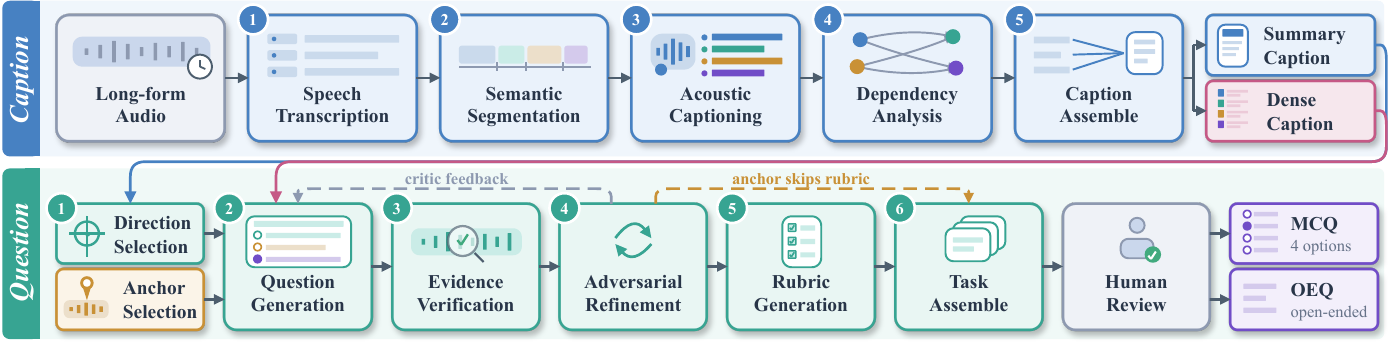}
\caption{\benchmark\ construction pipeline. Phase~1 (top) converts each long-form audio into a structured caption via a five-step pipeline. Phase~2 (bottom) generates questions under a shared six-step framework; Native QA runs all steps, while Anchor QA skips rubric generation.}
\label{fig:data-construction}
\end{figure}

Figure~\ref{fig:data-construction} shows the automated two-phase pipeline that constructs \benchmark. \textbf{Phase~1} converts each audio into a structured caption (\S\ref{sec:construction-caption}). \textbf{Phase~2} generates questions from that caption through two pipelines that share a common framework: the Native QA pipeline (\S\ref{sec:construction-native}) derives questions from the caption and verifies their ground truth against the audio, while the Anchor QA pipeline (\S\ref{sec:construction-anchor}) plants anchors into the audio and reads ground truth from the anchor settings. Quality assurance (\S\ref{sec:construction-quality}) runs during and after generation. Models used at each stage are listed in Appendix~\ref{app:models}.

\subsection{Caption Pipeline}
\label{sec:construction-caption}

{\setlength{\parskip}{1.5pt}%

The caption pipeline converts each audio into a structured caption in five steps.

\paragraph{1) Speech Transcription.}
An ASR model produces sentence-level transcripts with timestamps.

\paragraph{2) Semantic Segmentation.}
A language model splits the transcript into contiguous, semantically coherent segments using cues such as topic shifts, speaker changes, and narrative transitions, each capped at about 5 minutes and annotated with a title, summary, and key topics.

\paragraph{3) Acoustic Captioning.}
Each segment's audio is captioned by a multi-modal model, adding time-stamped acoustic details beyond the transcript: speaker identities and voice characteristics, paralinguistic cues, environment descriptions, and non-speech audio events.

\paragraph{4) Dependency Analysis.}
A language model extracts a speaker registry and cross-segment dependencies (e.g., causal links, contrasts, references), and produces global context (overall topic, genre, speaker roles).

\paragraph{5) Caption Assemble.}
Outputs are merged into the final caption and indexed as \emph{evidence records}, each a verifiable fact with a segment, time range, and modality tag, serving as the grounding units for question generation. Appendix~\ref{app:caption} details the assembled caption's two views.

}

\subsection{Native QA Pipeline}
\label{sec:construction-native}

{\setlength{\parskip}{1.5pt}%

The native QA pipeline derives questions from the audio's natural content, covering all three cognitive levels. A language model creates the questions across the first two steps.

\paragraph{1) Direction Selection.}
For each cognitive level, the language model chooses a free question direction and the segments relevant to it, drawing on the summary caption and the question space (Appendix~\ref{app:taxonomy}). The dense captions of those segments supply detail for question generation.

\paragraph{2) Question Generation.}
From the dense captions of the relevant context, the language model concretizes the direction into a multiple-choice candidate: a question stem, four options, the correct option, and supporting evidence records. Lessons from the critic feedback loop (\S\ref{sec:construction-quality}) are fed back into the prompt, so the model progressively avoids previously observed failure modes.

\paragraph{3) Evidence Verification.}
Cited evidence records are verified against the audio (\S\ref{sec:construction-quality}).

\paragraph{4) Adversarial Refinement.}
Surviving candidates pass through a three-level solver and a critic feedback loop that retain only audio-dependent items (\S\ref{sec:construction-quality}).

\paragraph{5) Rubric Generation.}
Each retained MCQ is first recast as an OEQ. From the solver's multiple reasoning paths, common and necessary points are distilled into essential/supporting rubric criteria, and a reference answer is synthesized.

\paragraph{6) Task Assemble.}
Items are unified to a common format, and options are shuffled to keep the correct answer balanced across positions, preventing position bias.

}

\subsection{Anchor QA Pipeline}
\label{sec:construction-anchor}

{\setlength{\parskip}{1.5pt}%

The anchor pipeline shares the native framework but differs in its first two steps. The other four are reused, with verification and refinement on U/R only and no rubric generation.

\paragraph{1) Anchor Selection.}
Guided by the summary caption, the pipeline selects 1--3 anchor positions that carry rich semantic context and relate to other segments. The anchor operation (Appendix~\ref{app:anchor}) is then applied, producing the modified audio and ground truth by construction.

\paragraph{2) Question Generation.}
From the anchor operations and their ground truth, together with the dense captions of the anchor context and its related segments and their known relations, the three-level question chain is generated.

}

\subsection{Quality Assurance}
\label{sec:construction-quality}

{\setlength{\parskip}{1.5pt}%

Quality assurance combines automatic gates and a critic feedback loop during generation with a human review afterward. Appendix~\ref{app:quality} details the gate settings, reports each gate's pass and rejection rates on a construction run, and describes the review procedure.

\paragraph{Evidence verification.}
Every question must cite evidence records from the caption. Each cited fragment is validated against the audio by a multi-modal model, and a candidate is kept only if all its records are confirmed, guarding the benchmark against caption errors.


\paragraph{Three-level solver.}
The solver is a language model that answers each candidate at three levels of access. The \emph{blind} level sees only the stem and options and rejects guessable items. The \emph{text-only} level adds the transcript and rejects items answerable from the spoken words alone. The \emph{full} level adds the dense caption, from which the model draws multiple reasoning paths and verifies solvability. A candidate is kept only if it fails the first two levels and the full level solves it.

\paragraph{Critic feedback loop.}
A critic agent analyzes how the blind and text-only solvers reached their answers, summarizing a root cause for each failure mode and the lessons shared across them. These lessons are kept in a memory bank and fed into later generation, so subsequent questions avoid the same shortcuts. The Native and Anchor pipelines keep separate memory banks.

\paragraph{Human review.}
Every item that clears the automatic gates receives a final human review against the audio, where annotators verify the question and its answer and, for anchor items, that the edit is audible and distinct from existing events, then accept it, apply a minor fix, or reject it.

}

\section{Experiments}
\label{sec:experiments}

\subsection{Experimental Setup}
\label{sec:exp-setup}

\paragraph{Audio models.}
We evaluate 12 models under a single native-audio protocol: each receives the audio itself, truncated to its input limit when it cannot ingest the whole clip, never a transcript in place of audio (App.~\ref{app:setup}). Seven are open-source, spanning audio specialists (Audio Flamingo Next, Voxtral-Mini and Voxtral-Small, MOSS-Audio) and omni-modal models (Phi-4-Multimodal, Baichuan-Omni-1.5, Qwen3-Omni); five are closed and all omni-modal, three Gemini versions (2.5-Pro, 3-Flash, 3.1-Pro) and two Qwen3.5-Omni tiers (Flash and Plus). Models that expose a thinking mode are run in both standard and thinking configurations. Per-model input limits and the specs behind them are given in App.~\ref{app:setup} (Table~\ref{tab:eval-limits}).

\paragraph{Text-only baselines.}
Three text-only baselines, all GPT-5.4 with no audio, bound the task at rising levels of textual access. \emph{Question-only} sees the question and its options alone, a floor set by answer-choice priors. \emph{Transcript} adds the ASR transcript, a rung reading only the spoken words. \emph{Caption} adds the dense authoring caption the questions were written from (for Chain, its anchor-augmented form), a text reference over the material the benchmark was built on.

\subsection{Overall Performance}
\label{sec:exp-overall}

We read Table~\ref{tab:main-results} along three axes: the overall level of performance, the gap between models, and the variation within each model.

\begin{table}[t]
\renewcommand{\yes}{\ding{51}}\renewcommand{\no}{\ding{55}}
\caption{Main results across three evaluation modes, \emph{Accuracy}
(Native-MCQ), \emph{Rubric} (Native-OEQ), and \emph{Chain} (Anchor-MCQ), and three
duration tiers (S/M/L); \emph{Avg} is the unweighted mean over the tiers. \textbf{Limit} is each model's audio-input cutoff in minutes
(\emph{--}: untruncated or not applicable) and \yes\ marks a thinking variant;
\textcolor{ltrunc}{gray} cells are scored on audio truncated to that limit
(protocol and per-model limits in \S\ref{sec:exp-setup}).
Best per column in \textbf{bold}, second-best \underline{underlined} (audio
models only).}
\label{tab:main-results}
\centering
\setlength{\tabcolsep}{5pt}
\renewcommand{\arraystretch}{1.2}
\fittable{%
\footnotesize
\begin{tabular}{l @{\hspace{6pt}} c @{\hspace{5pt}} c @{\hspace{5pt}} c *{3}{wc{0.6cm}} A *{3}{wc{0.6cm}} A *{3}{wc{0.6cm}} A}
\toprule
& & & & \multicolumn{4}{c}{\textbf{Accuracy (\%)}}
& \multicolumn{4}{c}{\textbf{Rubric (\%)}}
& \multicolumn{4}{c}{\textbf{Chain (\%)}} \\
\cmidrule(lr){5-8} \cmidrule(lr){9-12} \cmidrule(lr){13-16}
\textbf{Model} & \textbf{Size} & \textbf{Think} & \textbf{Limit} & \textbf{S} & \textbf{M} & \textbf{L} & \textbf{Avg} & \textbf{S} & \textbf{M} & \textbf{L} & \textbf{Avg} & \textbf{S} & \textbf{M} & \textbf{L} & \textbf{Avg} \\
\midrule
\multicolumn{16}{c}{\textbf{\textit{Text-only Baselines}}} \\
\midrule
GPT-5.4 (question-only) & -- & \yes & -- & 41.1 & 41.1 & 40.6 & 40.9 & -- & -- & -- & -- & 8.9 & 11.9 & 10.3 & 10.4 \\
\rowstripe GPT-5.4 (transcript) & -- & \yes & -- & 64.8 & 60.7 & 53.4 & 59.7 & 41.0 & 41.5 & 37.2 & 39.9 & 9.4 & 14.2 & 14.2 & 12.6 \\
GPT-5.4 (caption) & -- & \yes & -- & 83.2 & 80.0 & 77.3 & 80.2 & 59.1 & 57.4 & 53.0 & 56.5 & 66.7 & 64.4 & 53.1 & 61.4 \\
\midrule
\multicolumn{16}{c}{\textbf{\textit{Open-source Models}}} \\
\midrule
\rowstripe Phi-4-Multimodal & 5.6B & \no & 30m & 33.7 & \bl{34.6} & \bl{34.2} & 34.2 & 18.4 & \bl{13.6} & \bl{10.1} & 14.0 & 12.9 & \bl{7.2} & \bl{8.8} & 9.6 \\
& & \no & 30m & 46.1 & \bl{43.6} & \bl{37.2} & 42.3 & 16.1 & \bl{12.2} & \bl{13.7} & 14.0 & 5.3 & \bl{5.6} & \bl{6.1} & 5.7 \\
\multirow{-2}{*}{AudioFlamingoNext} & \multirow{-2}{*}{8B} & \yes & 30m & 43.3 & \bl{38.9} & \bl{35.0} & 39.1 & 26.9 & \bl{20.3} & \bl{16.3} & 21.2 & 7.5 & \bl{11.1} & \bl{13.1} & 10.6 \\
\rowstripe Voxtral-Mini & 3B & \no & 40m & 45.3 & \bl{35.8} & \bl{37.2} & 39.4 & 25.6 & \bl{21.8} & \bl{13.6} & 20.3 & 5.6 & \bl{9.4} & \bl{18.3} & 11.1 \\
Voxtral-Small & 24B & \no & 40m & 42.5 & \bl{40.0} & \bl{31.9} & 38.1 & 28.9 & \bl{24.2} & \bl{20.7} & 24.6 & 12.5 & \bl{4.7} & \bl{4.7} & 7.3 \\
\rowstripe & & \no & 50m & 45.0 & \bl{43.1} & \bl{35.6} & 41.2 & 23.7 & \bl{17.7} & \bl{10.7} & 17.4 & 8.3 & \bl{11.4} & \bl{9.2} & 9.6 \\
\rowstripe \multirow{-2}{*}{MOSS-Audio} & \multirow{-2}{*}{8B} & \yes & 50m & 52.2 & \bl{52.8} & \bl{38.6} & 47.9 & 30.1 & \bl{15.4} & \bl{11.6} & 19.0 & 13.9 & \bl{15.3} & \bl{11.4} & 13.5 \\
Baichuan-Omni-1.5 & 11B & \no & 80m & 46.9 & 46.9 & \bl{36.4} & 43.4 & 19.0 & 12.5 & \bl{9.7} & 13.7 & 11.9 & 14.4 & \bl{14.7} & 13.7 \\
\rowstripe & & \no & 80m & 53.9 & 54.4 & \bl{46.2} & 51.5 & 31.3 & 24.3 & \bl{20.6} & 25.4 & 20.5 & 19.2 & \bl{13.2} & 17.6 \\
\rowstripe \multirow{-2}{*}{Qwen3-Omni} & \multirow{-2}{*}{\makecell{30B\\(A3B)}} & \yes & 80m & 56.9 & 57.5 & \bl{49.2} & 54.5 & 38.4 & 29.2 & \bl{22.8} & 30.1 & 7.8 & 10.6 & \bl{15.3} & 11.2 \\
\midrule
\multicolumn{16}{c}{\textbf{\textit{Closed-source Models}}} \\
\midrule
Gemini-2.5-Pro & -- & \yes & -- & 67.5 & 64.2 & 54.7 & 62.1 & 46.9 & 40.7 & 36.4 & 41.3 & \second{24.2} & \second{25.3} & \second{21.9} & \second{23.8} \\
\rowstripe Gemini-3-Flash & -- & \no & -- & 65.6 & 69.2 & 64.2 & 66.3 & 54.9 & \best{55.0} & \best{53.8} & \best{54.6} & 16.7 & 20.6 & 11.9 & 16.4 \\
Gemini-3.1-Pro & -- & \yes & -- & \second{76.7} & \best{74.4} & \second{65.8} & \second{72.3} & \second{56.2} & \second{48.8} & 43.3 & 49.4 & 17.5 & \second{25.3} & 20.0 & 20.9 \\
\rowstripe Qwen3.5-Omni-Flash & -- & \no & -- & 73.9 & \second{71.4} & 58.9 & 68.1 & 50.3 & 40.9 & 35.9 & 42.4 & 23.1 & 19.2 & 18.3 & 20.2 \\
Qwen3.5-Omni-Plus & -- & \no & -- & \best{79.2} & \best{74.4} & \best{70.3} & \best{74.6} & \best{56.9} & 48.6 & \second{47.0} & \second{50.8} & \best{30.8} & \best{33.1} & \best{23.1} & \best{29.0} \\
\bottomrule
\end{tabular}%
}
\end{table}

\textbf{Current models are burdened by the audio itself, a long and redundant signal they must distill.} Handed the same content as text, the dense authoring caption the questions were written from rather than the audio, a text-only solver (GPT-5.4, caption baseline) leads every audio model on Accuracy and Rubric. What separates the two is modality: one side reads the facts off text, the other must recover them from the waveform. The text side winning places the difficulty before reasoning, in gathering usable information from the audio: an audio encoder emits tens of frames per second, so the model must sift a long and largely redundant stream for the few facts a question turns on.

\textbf{Long-form audio exposes a clear gap between closed- and open-source models.} The open models are also limited in reach: each is capped at a $30$--$80$\,min input (grey cells) and sees only part of the Medium and Long audio, while the closed models run untruncated. The gap holds even where reach is equal: on the Short tier, which every model ingests in full, every closed model outscores every open model on Accuracy, and the same separation holds on Rubric. The exception is Chain, whose scoring is bottom-loaded (Figure~\ref{fig:chain-baseline}): with every model compressed toward the floor, the closed and open ranges overlap, though the highest scores stay closed.

\textbf{Performance declines steadily with duration, worst on hour-long inputs.} The fall spans all three modes: Accuracy, Rubric, and Chain each drop from S to L, and the L tier (avg.\ $\sim$90\,min) is the lowest for nearly every configuration. Part of this is length alone: the text-only transcript and dense-caption baselines lose $6$--$11$ accuracy points from S to L, while the content-free question-only baseline holds flat. Longer inputs thus tax comprehension even as clean text, a long-context penalty the audio models carry on top of distilling the signal.

Two cross-cutting factors round out the picture. Enabling \emph{thinking} generally helps, most clearly on open-ended Rubric. \emph{Language} does not drive the ordering: the strongest models are language-balanced, and a sizable gap appears only on the weaker open models (App.~\ref{app:language}).

\subsection{Cognitive levels}
\label{sec:exp-native}

\begin{figure}[t]
\centering
\setlength{\abovecaptionskip}{4pt}
\begin{subfigure}[t]{0.49\textwidth}
\centering
\includegraphics[width=\linewidth]{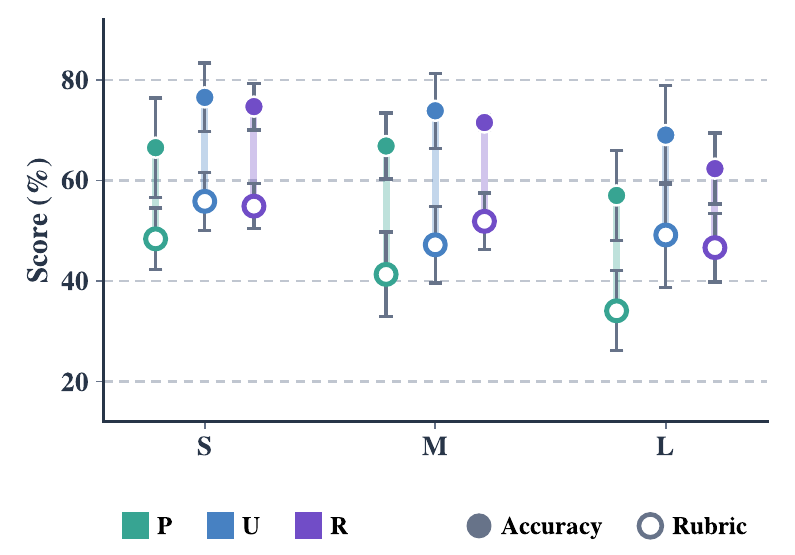}
\caption{Native QA}
\label{fig:native-qa}
\end{subfigure}
\hfill
\begin{subfigure}[t]{0.49\textwidth}
\centering
\includegraphics[width=\linewidth]{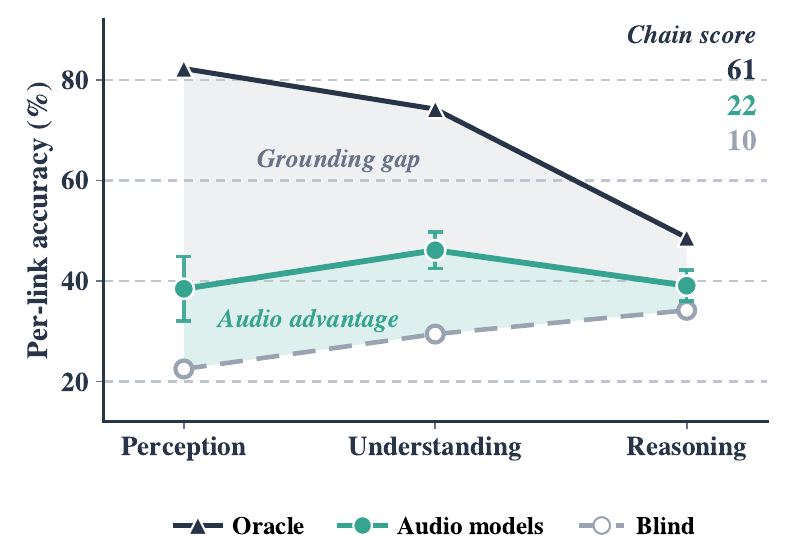}
\caption{Anchor QA}
\label{fig:anchor-chain}
\end{subfigure}
\caption{Scores by cognitive level (Perception, Understanding, Reasoning). (\subref{fig:native-qa})~Native QA per duration tier: filled dots Accuracy, open Rubric, ticks $\pm 1$ SD. (\subref{fig:anchor-chain})~Anchor QA per-link accuracy for a blind solver, the audio models ($\pm 1$ SD), and a text oracle given the anchors; corner ladder is the truncated chain score (Eq.~\ref{eq:chain}).}
\label{fig:levels-combined}
\end{figure}

Figure~\ref{fig:levels-combined} splits both question paths by cognitive level.

\textbf{Perception is the weakest cognitive level.} Across the closed-source models it is the lowest level on both paths, and its weakness shows most once scoring moves past a multiple-choice pick. Under both metrics perception ranks lowest, but the drop is far steeper under Rubric and widens with duration: on the M and L tiers, perception sits well below understanding and reasoning (Figure~\ref{fig:native-qa}). The deficit is specific: recognition of content keeps pace with the higher levels, while time-sensitive perception lags behind. Within perception, localization and counting are the weak dimensions, and they fail differently (App.~\ref{app:tasktype}): counting is low under both formats alike, a capacity limit no scoring change lifts, while localization collapses only once the options are withheld. Anchor QA isolates the same cost: hand a text oracle the anchor positions and its perceptual link clears the audio models by the widest margin anywhere on the chain, a gap that closes toward reasoning only because the oracle itself falls there (Figure~\ref{fig:anchor-chain}).

\textbf{Grounding an answer in the audio is the core failure.} It surfaces once scoring makes a model produce a grounded answer, on both question paths. On the native path, a correct multiple-choice pick only weakly predicts a good open-ended answer: matched item by item on the closed-source models, localization keeps only about half its accuracy once the options are withheld (App.~\ref{app:mcq-oeq}), because four options let a model confirm a timestamp it cannot itself produce. In the open-ended answers, models describe what happens far more reliably than when it happens, even on reasoning questions (App.~\ref{app:content-timing}). On the anchor path, grounding becomes a requirement: a link earns credit only when the model has named the anchor, so the truncated chain score (Eq.~\ref{eq:chain}) falls far below per-link accuracy, for audio models and a blind, no-audio solver alike. Even on anchors it never located, a model answers understanding nearly as often as when it did (App.~\ref{app:anchor-analysis}): the higher links are answered about as much by guessing as by grounding. The effect is sharpest where the anchor is hardest to find: reasoning peaks on silenced spans, the operation whose perception link is weakest, while their chain score stays the lowest of the three.

\section{Conclusion}
\label{sec:conclusion}

\benchmark\ is a benchmark for grounded, long-form audio comprehension, pairing durations from minutes to hours with three levels of cognitive depth, each scored through matched multiple-choice and open-ended formats. Across 12 LALMs, the failure is consistent: before reasoning is even tested, models cannot reliably ground what they describe in the audio itself, and the gap widens as audio lengthens. \benchmark\ offers a baseline for tracking progress as models learn to sustain comprehension across longer spans and close this gap. The relevant measure is less the amount of audio a model can accept than the amount it can use.

\section*{AI Usage Statement}
In this work, we used generative AI tools as components of our method and for manuscript preparation. Large language model APIs are part of \benchmark's automated construction pipeline, generating structured captions, questions, and adversarial critic feedback (\S\ref{sec:construction}, Appendix~\ref{app:sec-construction}), and an LLM judge scores open-ended responses (\S\ref{sec:design-taxonomy}, Appendix~\ref{app:sec-eval}). All pipeline output passes the verification and human review steps described there before entering the benchmark. We additionally used AI writing assistants to polish the manuscript's prose. We did not use generative AI to develop theoretical models, formulate or prove mathematical claims, propose or refine the research hypotheses, or design the research methodology. We have reviewed all AI-assisted work, and we take responsibility for the final content of this work, including text, claims, and artifacts produced with the aid of generative AI.

\section*{Reproducibility Statement}
The construction pipeline (captioning, QA generation, adversarial critic feedback) is described in Appendix~\ref{app:sec-construction}. The evaluation protocol, scoring rubrics, and judge prompts are in Appendix~\ref{app:sec-eval}, and the models and API versions used are listed in Appendix~\ref{app:models}. \benchmark\ is available at the anonymized link in the abstract during review; the real repository will be made public on acceptance.

\bibliography{iclr2027_conference}
\bibliographystyle{iclr2027_conference}

\clearpage
\appendix

\etocsettocdepth.toc{subsection}
\begingroup
\renewcommand{\contentsname}{Appendix Contents}
\etocsetnexttocdepth{subsection}
\tableofcontents
\endgroup
\bigskip

\section{Data Ethics}
\label{app:sec-ethics}

\subsection{Licensing and Release}
\label{app:licensing}

\benchmark\ is released for non-commercial research and evaluation. The artifacts we produce, QA items, rubric criteria, anchor edit manifests, sound events, and evaluation scripts, are distributed under a CC~BY-NC-SA~4.0 license. The source recordings are publicly available media collected from the Internet, and copyright of each recording remains with its original creator. The recordings are included only for evaluation, and downloading the dataset constitutes agreement not to redistribute the audio or use it in commercial products.

\subsection{Ethics and Responsible Use}
\label{app:ethics}

All audio comes from publicly available media; speakers are referenced by role rather than identity, and no personal information in the audio is used as an answer key. \benchmark\ is an evaluation resource and is not intended for training speaker-identification or surveillance systems; the anchor perturbations exist only to probe grounding. Human reviewers were compensated at or above local market rates and worked only with publicly available media.

\section{Construction Process}
\label{app:sec-construction}

\subsection{Caption Pipeline}
\label{app:caption}

Every question is written from a structured caption, not from the raw audio (\S\ref{sec:construction}). The caption is read in two views. The \emph{summary} view is an audio-level overview: global topic, narrative arc, and tone; a speaker registry; a segment index; and the cross-segment dependencies between them. It drives direction selection for Native QA and anchor selection for Anchor QA, the stages that reason over the whole audio at once. The \emph{dense} view holds per-segment detail: the acoustic environment, a timestamped transcript with paralinguistic annotation, and discrete audio events. Generation reads the global context together with the dense captions of only the segments a question targets.

Both views are grounded in \emph{evidence records}. Assembling the caption flattens it into one record per groundable unit, each carrying a stable ID, its segment, a time range, a modality (speech, sound, environment, cross-segment, or global), and the grounded content. The generator cites the IDs it relied on, and verification re-checks exactly those IDs against the audio before a candidate is kept (\S\ref{sec:construction-quality}). These records are the grounding unit referenced throughout \S\ref{sec:construction}. Figures~\ref{fig:cap-summary} and~\ref{fig:cap-dense} render one audio sample through both views.

\begin{figure}[t]
\begin{casebox}{Summary Caption}
\begin{alltt}
\cbhd{GLOBAL}
  topic:  how status and power operate in professional life, and how respect and
          influence are built.
  arc:    defines status as perception-based, distinct from power as resource control;
          reframes both as legitimate, then moves into practical frameworks. ...
  tone:   insightful, pragmatic, encouraging; warm.

\cbhd{SPEAKERS}
  spk_1  male host     guides conversation, asks questions
  spk_2  female guest  primary expert on status and power
  spk_3  male guest    secondary participant

\cbhd{SEGMENT}  (id . time . event_type . title)
  seg_000  0-26s      monologue   Status as a construct shaped by others' judgments
  seg_001  26-157s    interview   The guest's path into power and status research
  seg_002  157-310s   discussion  Defining power and status and their difference
  ...
  seg_031  3731-3759s transition  Final acknowledgments and goodbye

\cbhd{DEPENDENCIES}  (source -> target . type . relation)
  seg_000 -> seg_002   definition refinement
     the opening claim that status lives in others' minds is formalized in seg_002 via
     the contrast with power as control over resources.
  seg_002 -> seg_004   causal
     after both are defined, seg_004 explains why they matter. ...
\end{alltt}
\end{casebox}
\caption{Summary caption: an audio-level overview (global topic, narrative arc, and tone; a speaker registry; a segment index; and cross-segment dependencies). It drives direction selection for Native QA and anchor selection for Anchor QA.}
\label{fig:cap-summary}
\end{figure}

\begin{figure}[t]
\begin{casebox}{Dense Caption}
\begin{alltt}
\cbhd{SEGMENT}  seg_001 . 25.7-156.7s . interview

\cbhd{ENVIRONMENT}
  Quiet professional studio; no reverb or background noise.

\cbhd{SPEECH}
  30.0-31.2s    spk_1: "All right welcome to the show"
                (warm, welcoming, slight rising intonation)
  32.3-33.0s    spk_2: "Happy to be here"  (brief, cheerful, upbeat)
  34.0-41.4s    spk_1: "I'm really curious what drew you to organizational psychology,
                and to power and status in the workplace"
                (measured, thoughtful, curious)
  ...           (continues to 156.6s)

\cbhd{AUDIO EVENTS}
  93.6-93.8s    quiet mouth click / lip smack
\end{alltt}
\end{casebox}
\caption{Dense caption: per-segment detail (acoustic environment, a timestamped transcript with paralinguistic annotation, and discrete audio events), shown for one segment. Every segment carries the same fields, and generation reads the global context together with the dense captions of the segments a question targets.}
\label{fig:cap-dense}
\end{figure}

\subsection{Native QA Pipeline}
\label{app:taxonomy}

Table~\ref{tab:question-directions} lists question directions for each (cognitive level, dimension) pair introduced in \S\ref{sec:design-taxonomy}. Following the free scheme of \S\ref{sec:design-taxonomy}, the generator treats these directions as inspiration, deciding the actual direction of each question for the audio at hand and combining several where useful (\S\ref{sec:construction-native}).

\begin{table}[t]
\centering
\setlength{\tabcolsep}{5pt}
\renewcommand{\arraystretch}{1.25}
\caption{Native QA question directions per cognitive level and dimension.}
\label{tab:question-directions}
\fittable{%
\footnotesize
\begin{tabular}{p{1.9cm} @{\hspace{16pt}} p{2.2cm} p{11.5cm}}
\toprule
\textbf{Level} & \textbf{Dimension} & \textbf{Question Space} \\
\midrule
\multirow{3}{1.9cm}{\textbf{Perception}}
 & Recognition & Linguistic recognition; Speaker identification; Sound scene classification; Subtle audio discrimination; Prosodic and paralinguistic recognition; Audio source discrimination; Overlapping speech resolution; Sound event recognition; Music recognition \\[3pt]
 & \cellcolor{lstripebg}Localization & \cellcolor{lstripebg}Keyword/phrase localization; Speaker turn localization; Sound event localization; Temporal ordering; Relative temporal localization; Duration estimation \\[3pt]
 & Counting & Linguistic counting; Speaker counting; Sound event counting; Conditional counting; Comparative counting; Frequency/density judgment \\
\midrule
\multirow{3}{1.9cm}{\textbf{Understanding}}
 & Interpretation & Role/relationship recognition; Emotional atmosphere; Epistemic stance recognition; Subtext and implicit intent; Rhetorical device recognition; Social function recognition; Silence and non-response semantics; Content structure recognition; Quotation and paraphrase recognition; Humor and comedic intent recognition \\[3pt]
 & \cellcolor{lstripebg}Tracking & \cellcolor{lstripebg}Topic transition detection; Emotional/stance evolution; Narrative progression; Engagement/dominance shift; Discourse function shift; Information density change; Structural segment transition; Prosodic consistency tracking \\[3pt]
 & Summarization & Holistic summarization; Key point extraction; Speaker contribution analysis; Surface vs.\ actual consensus; Selective summarization; Omission detection; Bias detection; Segment importance judgment \\
\midrule
\multirow{3}{1.9cm}{\textbf{Reasoning}}
 & Causal & Attribution; Foreshadowing; Causal chain reconstruction; Competing explanations; Correlation vs.\ causation; Prosodic causation \\[3pt]
 & \cellcolor{lstripebg}Relational & \cellcolor{lstripebg}Long-range reference resolution; Cross-segment information linking; Retrospective reinterpretation; Structural segment contextual relevance \\[3pt]
 & Logical & Contradiction detection; Stance reversal detection; Argument validity assessment; Information completeness check; Audio atmosphere vs.\ stated content divergence \\
\bottomrule
\end{tabular}%
}
\end{table}

\subsection{Anchor QA Pipeline}
\label{app:anchor}

Anchor QA uses one operation of each type, applied to the audio before any question is written (\S\ref{sec:construction-anchor}): addition, deletion, and modification. Each leaves a trace that is localizable to a point in time and layered across the perception, understanding, and reasoning levels, and each is applied at one to three positions so the perception question can ask how many times the trace occurs and where a given one falls.

\emph{Addition} mixes an external sound event into the audio at the chosen positions. Across the anchor set these events span 62 distinct categories, from animal calls and alarms to mechanical impacts and ambient nature. The event clips are synthesized with Stable Audio 3~\citep{evans2026stableaudio3}: each category is rendered from a short text prompt, so the clips stay consistent within a category and carry no stock-audio fingerprint a model could have memorized. The event is laid over the existing audio at a controlled signal-to-noise ratio, roughly 6--10\,dB, and for each sample any category that could already occur in it is excluded, so the added event cannot be mistaken for one that was there to begin with. A single event is reused across positions, so the chain can ask how often it occurs and where.

\emph{Deletion} obscures a chosen span so its content can no longer be made out, while the span keeps its place and length on the timeline. The span is either silenced outright or smeared into unintelligibility by a low-pass filter near 300\,Hz mixed with a modulated noise bed. What the listener notices is that something at that point has dropped out, which is the cue the perception question tests.

\emph{Modification} alters an acoustic attribute over a span while leaving the words spoken there unchanged: either loudness, raised or lowered by 8--12\,dB, or playback speed, sped up or slowed by a factor of 1.5--3 through time-stretching. The delivery changes, so the trace is a shift in how a passage sounds.

\emph{Anchor selection} reads the summary view in two passes. The first shortlists a few candidate segments, each picked for its semantic weight and for the relations it holds to other segments, which later give the reasoning question somewhere to point. The second fixes a precise window inside a chosen segment, held within that segment, kept clear of the audio's opening and close, floored at a few seconds and capped to a small fraction of the total so the change stays local. The window is placed over active speech, so the operation lands on audible content instead of being lost in a pause. When an operation is placed at several positions, they are spread across distinct segments, keeping each occurrence separately locatable. These placement constraints, an added event that is not already present and a position clear of silence, are verified once more during human review (\S\ref{app:quality}).

\emph{Chain construction} builds the three questions so that their dependency lives in the options: each reads on its own, yet the wrong options are laid out so an error at one level leads to a consistent one at the next (\S\ref{sec:design-taxonomy}). The perception question crosses two axes in a single item, pairing a correct or off-by-one count of the trace with either its true position or a far-off point; asked together, count and position cannot be settled one without the other. The understanding question offers the real content found at each position, so a miscount that reindexes ``the Nth'' occurrence lands on another position's genuine material. The reasoning question offers cross-segment relations drawn from those same positions, plus a ``no clear relation'' foil, so a position carried over wrong selects a true but misattributed relation. The perception options are assembled directly from the recorded ground truth; the understanding and reasoning options are written by the generator from the captioned anchor context (\S\ref{sec:construction-anchor}).

\begin{wrapfigure}{r}{0.45\textwidth}
\vspace{-1.5\baselineskip}
\centering
\setlength{\abovecaptionskip}{2pt}
\includegraphics[width=\linewidth]{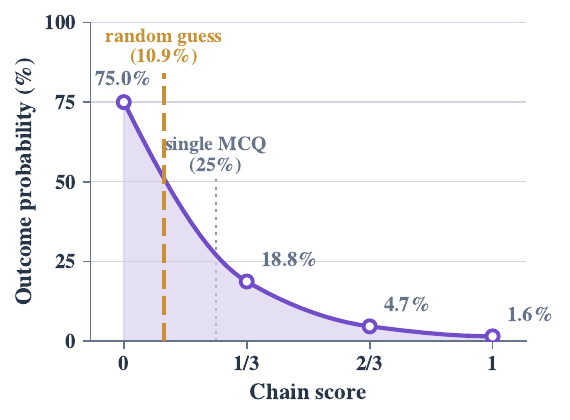}
\caption{Anchor \emph{Chain} score under random guessing. The four possible outcomes $\{0, 1/3, 2/3, 1\}$ are marked; the curve interpolates their probabilities.}
\label{fig:chain-baseline}
\vspace{-1.2\baselineskip}
\end{wrapfigure}

\emph{Random baseline.} The three chained questions are each four-option multiple-choice items, so an uninformed guess is correct with probability $1/4$ at every level, independently. Since the chain credits only the longest correct prefix (Eq.~\ref{eq:chain}), a random responder scores $0$ unless it first clears the perception question, and earns $k/3$ for the first $k$ levels it happens to clear. The induced distribution is bottom-loaded: it lands on $0$ three quarters of the time and reaches a full $1$ only once in sixty-four attempts (Figure~\ref{fig:chain-baseline}), for an expected score of $10.9\%$. That floor is below the $25\%$ of a single four-option MCQ precisely because a late correct answer earns nothing once an earlier one is wrong, which is the property that makes the chain resistant to guessing.

\subsection{Quality Assurance}
\label{app:quality}

\begin{wraptable}{r}{0.42\textwidth}
\vspace{-1.5\baselineskip}
\caption{Automatic-gate yields (\%) by cognitive level (P/U/R) and overall; \emph{Solver rejection} rows are shares of solver inputs, the rest pass.}
\label{tab:qa-yield}
\small
\setlength{\tabcolsep}{3pt}
\renewcommand{\arraystretch}{1.1}
\begin{tabular*}{\linewidth}{@{\extracolsep{\fill}} l cccc @{}}
\toprule
\textbf{(\%)} & \textbf{P} & \textbf{U} & \textbf{R} & \textbf{All} \\
\midrule
Evidence pass & 69 & 90 & 88 & 88 \\
\addlinespace[2pt]
\rowcolor{lheaderbg} \multicolumn{5}{@{}l}{\textit{Solver rejection}} \\
\quad Blind & 9 & 32 & 41 & 37 \\
\quad Text-only & 21 & 48 & 37 & 41 \\
\quad Unsolvable & 19 & 6 & 16 & 12 \\
\midrule
Survival & 29 & 11 & 5 & 8 \\
\bottomrule
\end{tabular*}
\end{wraptable}

A few settings behind the automatic gates of \S\ref{sec:construction-quality} are worth stating. Each solver level answers the candidate over three independent rollouts, and the gate compares how often it succeeds against a threshold, so a pass reflects a real shortcut rather than a single lucky draw: an item is discarded if the blind solver is correct in more than $1/3$ of its rollouts, or the text-only solver in more than $2/3$. Blind is the stricter bar because the stem and options alone should almost never suffice, whereas the transcript legitimately carries content and is disqualifying only when it settles the answer on nearly every rollout. The full solver must itself solve the item, confirming that it is answerable once the audio is in hand. When a gate flags an item, the critic names the shortcut it exploited from a fixed catalog, among them specificity bias, semantic leakage, elimination by absurdity, common-sense giveaway, and transcript sufficiency, and stores the lesson under that name so later generation is steered away from the specific pattern rather than by a generic warning. Figure~\ref{fig:critic-memory} shows a few examples from the resulting bank, grouped by cognitive level.

The gates are the dominant filter on candidate quality. Table~\ref{tab:qa-yield} reports their pass and rejection rates on a construction run: evidence verification and the solver together retain roughly $8\%$ of candidates, and most of what the solver rejects is shortcut-solvable, removed by the blind or text-only level. The three cognitive levels fail at different gates: perception items are most often caught at evidence verification for citing events absent from the audio, understanding items at the text-only level for being answerable from the transcript, and reasoning items at the blind level for being guessable from priors, which leaves reasoning the lowest survival rate. The memory bank's effect is visible within the run: candidates generated after the bank has accumulated lessons clear the solver gate more often than early ones, with the solver pass rate rising from roughly $8\%$ to $12\%$ and end-to-end survival from roughly $7\%$ to $9\%$, while the evidence-verification rate barely moves. The gain sits at the solver gate the bank targets, consistent with its lessons steering generation away from the shortcuts that gate catches.

\begin{figure}[t]
\begin{casebox}[gray]{Critic Memory Bank}
\begin{alltt}
\cbhd{PERCEPTION}
  Make questions genuinely audio-dependent: if the transcript already states the
  answer, paraphrases the intent, or directly explains the cause, the item is
  text-solvable and invalid. Target judgments that require tone, delivery, timing,
  emphasis, or cross-segment integration, and use distractors that are all
  plausible, transcript-grounded alternatives rather than obvious mismatches or
  unsupported negatives.

\cbhd{UNDERSTANDING}
  For shift- or transition-tracking questions, do not make the correct answer stand
  out by mirroring the stem, explicitly naming the target state, or using a
  different type of cue than the distractors. All options should be parallel,
  neutral timeline anchors from the same local context and level of abstraction, so
  the solver must identify the transition from the source's actual sequence rather
  than from lexical or structural giveaways.

\cbhd{REASONING}
  In causal multiple-choice questions, avoid making the correct answer the only
  option that echoes the stem's core contrast or wording. Instead, make all options
  plausible competing causes for the same outcome, matched in semantic family,
  relevance, and abstraction level, so the item tests inference rather than wording
  alignment.
\end{alltt}
\end{casebox}
\caption{Examples from the critic's memory bank, grouped by cognitive level. Text is quoted verbatim from the stored lessons.}
\label{fig:critic-memory}
\end{figure}

Human review runs item by item against the audio. The reviewer sees the open-ended question with its reference answer and rubric criteria, and the multiple-choice stem with its options and marked answer; timestamps in the reference answer and criteria point to the stretch of audio the item is about. The reviewer plays that stretch, checks the open-ended item first and then the multiple-choice one, and either accepts the item, applies a minor fix such as tightening a stem, an option, or a criterion, or rejects it when a factual error leaves it beyond repair. For anchor items the reviewer also plays the edited audio to confirm the anchor is audible, matches its stated operation, and cannot be confused with another event already in the audio.

\section{Evaluation Protocol}
\label{app:sec-eval}

\subsection{Evaluation Setup}
\label{app:setup}

\benchmark{} is run under a single input protocol: native audio only. A model that cannot ingest the whole audio still receives audio, truncated to its limit, never a text transcript as a fallback, so every score measures audio understanding rather than a cascade. Each model carries one \emph{max processing duration}, the \textbf{Limit} column of Table~\ref{tab:main-results}, and for an audio sample it receives $\min(\text{limit}, \text{length})$ as the leading prefix, taken content-agnostically rather than centered on where the answer lies.

Alongside the audio, each item carries a fixed one-line instruction prepended to the question, set by answer format. Multiple-choice items (Accuracy and Chain) constrain the reply to a single option letter, matched exactly against the key; open-ended items (Rubric) instead ask for audio timestamps, the citations that temporal\% scores, though omitting one is not penalized (App.~\ref{app:rubric}). The two templates:

\begin{promptbox}{Model Prompt}
@@MCQ
Listen carefully to the audio and answer the following question. Answer with ONLY the option letter.

@@OEQ
Listen carefully to the audio and answer the following question. Cite audio timestamps in your analysis.
\end{promptbox}

The limit follows a fixed priority order: a stated audio ceiling in the model card or technical report is taken directly; otherwise the limit is the largest audio length that fits the model's trained context, computed from the context window, the reserved text budget, and the audio encoding rate. The context is the trained range alone. The applied value is the highest length that runs cleanly under this ceiling, dropping where output degrades earlier (MOSS-Audio produces gibberish past roughly $50$\,min). Closed models run through APIs whose per-request duration caps clear the longest tier, so those models run untruncated. Table~\ref{tab:eval-limits} collects the per-model figures.

\begin{table}[t]
\centering
\footnotesize
\caption{Audio-input limit per evaluated model and the specs behind it, expanding the \textbf{Limit} column of Table~\ref{tab:main-results}. \textbf{Context} is the trained position range; \textbf{Tokens/s} is positions per second into the language model after audio pooling; \textbf{Limit} is the applied truncation cutoff, with \emph{--} for a model that runs untruncated on every tier.}
\label{tab:eval-limits}
\setlength{\tabcolsep}{10pt}
\renewcommand{\arraystretch}{1.25}
\fittable{%
\begin{tabular}{@{}l r c l c@{}}
\toprule
\textbf{Model} & \textbf{Context} & \textbf{Tokens/s} & \textbf{Basis for limit} & \textbf{Limit} \\
\midrule
\rowcolor{lheaderbg}\multicolumn{5}{@{}l}{\textbf{\textit{Open-source Models}}} \\
Phi-4-Multimodal \citep{abouelenin2025phi}      & 131{,}072 & --   & Reported ceiling & 30\,min \\
AudioFlamingoNext \citep{ghosh2026audio}        & 131{,}072 & --   & Reported ceiling & 30\,min \\
Voxtral-Mini \citep{mistral2025voxtral}         & 32{,}768  & 12.5 & Reported ceiling & 40\,min \\
Voxtral-Small \citep{mistral2025voxtral}        & 32{,}768  & 12.5 & Reported ceiling & 40\,min \\
MOSS-Audio \citep{openmoss2026moss}             & 40{,}960  & 13   & Position budget  & 50\,min \\
Baichuan-Omni-1.5 \citep{li2025baichuan}        & 65{,}536  & 12.5 & Position budget  & 80\,min \\
Qwen3-Omni \citep{xu2025qwen1}                  & 65{,}536  & 13   & Position budget  & 80\,min \\
\midrule
\rowcolor{lheaderbg}\multicolumn{5}{@{}l}{\textbf{\textit{Closed-source Models}}} \\
Gemini-2.5-Pro     & API & -- & Per-request cap 9.5\,h & -- \\
Gemini-3-Flash     & API & -- & Per-request cap 9.5\,h & -- \\
Gemini-3.1-Pro     & API & -- & Per-request cap 8.4\,h & -- \\
Qwen3.5-Omni-Flash & API & -- & Per-request cap 3\,h   & -- \\
Qwen3.5-Omni-Plus  & API & -- & Per-request cap 3\,h   & -- \\
\bottomrule
\end{tabular}%
}
\end{table}

\subsection{Scoring Details}
\label{app:rubric}

\begin{wraptable}{r}{0.46\textwidth}
\vspace{-2.5\baselineskip}
\centering
\footnotesize
\caption{Supporting-criterion score $s_c=\mathbf{1}[\sigma_c{>}0]\,(\sigma_c{+}\tau_c)/2$ by semantic verdict $\sigma_c$ and temporal verdict $\tau_c$.}
\label{tab:score-matrix}
\begin{tabular}{@{}lccc@{}}
\toprule
& \multicolumn{3}{c}{\textbf{Temporal} $\tau_c$} \\
\cmidrule(lr){2-4}
\textbf{Semantic} $\sigma_c$ & correct & wrong & none \\
\midrule
correct ($1$)   & $1.0$  & $0.5$  & $1.0$ \\
partial ($0.5$) & $0.75$ & $0.25$ & $0.5$ \\
wrong ($0$)     & $0$    & $0$    & $0$ \\
\bottomrule
\end{tabular}
\end{wraptable}

Each criterion $c$ carries an importance weight $w_c$ ($1$ for essential, $0.5$ for supporting) and is graded by the judge ensemble on two dimensions. The \emph{semantic} score $\sigma_c \in \{0, 0.5, 1\}$ marks the required content as wrong, partial, or correct. The \emph{temporal} score $\tau_c \in \{0, 1\}$ marks whether a predicted timestamp falls within the criterion's reference time range under a $\pm 5$\,s tolerance; timestamps in any format are normalized to seconds before comparison.

Which dimensions apply follows Eq.~\ref{eq:rubric}. An essential criterion is graded on one dimension: $\tau_c$ for a localization criterion, which asks when something occurs, and $\sigma_c$ otherwise. A supporting criterion takes the gated mean of $\sigma_c$ and $\tau_c$ when it carries a reference time range, and $\sigma_c$ alone otherwise (Table~\ref{tab:score-matrix}). Two conventions complete the rule: a semantically wrong criterion scores $s_c = 0$ regardless of timing, since a timestamp cannot ground a false statement; and a missing timestamp falls back to $\sigma_c$ rather than incurring a penalty, because answers are not required to cite time.

Because $\sigma_c$ and $\tau_c$ are logged separately, we read the rubric along each axis for the analysis of App.~\ref{app:content-timing} as two per-criterion sub-scores, \emph{semantic\%} and \emph{temporal\%}. Write $N_\sigma(\cdot)$ and $N_\tau(\cdot)$ for counts of semantic and temporal verdicts: a semantic verdict is \emph{correct}, \emph{partial}, or \emph{wrong} when a claim is demanded and \emph{none} on a pure-localization criterion; a temporal verdict is \emph{correct} or \emph{wrong} when a timestamp is given, \emph{missing} when one is due but omitted, and \emph{none} when no timing is demanded. Semantic\% is the mean semantic score over claim-demanding criteria, and temporal\% is the accuracy over attempted timestamps:
\begin{equation}
\mathrm{semantic\%}=\frac{N_\sigma(\mathrm{cor})+\tfrac12 N_\sigma(\mathrm{par})}{N_\sigma(\mathrm{cor})+N_\sigma(\mathrm{par})+N_\sigma(\mathrm{wro})}, \qquad
\mathrm{temporal\%}=\frac{N_\tau(\mathrm{cor})}{N_\tau(\mathrm{cor})+N_\tau(\mathrm{wro})}.
\end{equation}
Semantic\% excludes \emph{none} (no claim demanded); temporal\% excludes both \emph{none} (no timing demanded) and \emph{missing} (a timestamp due but omitted), so it scores localization conditional on an attempt and separates it from both content and silence.

Temporal\% is conditioned to match how each criterion type is scored. On an essential localization criterion the timestamp \emph{is} the answer ($\sigma_c$ is \emph{none}), so temporal\% is unconditioned. On a supporting criterion the timestamp is a citation and the gate $\mathbf{1}[\sigma_c>0]$ zeros the whole criterion when the claim is wrong, so timing bears on the score only when the claim is right; supporting temporal\% therefore counts $N_\tau$ over the rows where $\sigma_c$ is \emph{correct} alone.

The two sub-scores compose back into the rubric differently by importance. An essential criterion is graded on one axis, so its mean score is the share-weighted average $p_\sigma\,\mathrm{semantic\%}+p_\tau\,\mathrm{temporal\%}$, where $p_\sigma,p_\tau$ are the fractions of essential criteria graded on the semantic and temporal axis (about $0.73$ and $0.27$); this is a partition up to the ${\approx}0.2\%$ of essential criteria the judge grades on both. A supporting criterion is gated and dual-axis: semantic\% drives its score and temporal\% enters only as the halving penalty of Eq.~\ref{eq:rubric} on the claims it grades correct, so the essential--supporting gap of App.~\ref{app:content-timing} is the semantic--temporal gap re-expressed through criterion importance. App.~\ref{app:content-timing} reads this semantic vs.\ temporal split model by model and by importance.

\subsection{Judge Prompt}
\label{app:judge-prompt}

Each judge in the ensemble receives the system prompt below, followed by the question, the criterion list (each with \texttt{text}, \texttt{importance}, and an optional \texttt{time\_range}), and the model's response, and returns one \texttt{semantic}/\texttt{temporal} verdict pair per criterion.

\begin{promptbox}{Judge System Prompt}
You are an audio-understanding evaluation expert. For each evaluation criterion
provided, judge the quality of the model's response independently.

## Evaluation Dimensions

### Dimension Selection Rules

Choose which dimension(s) to evaluate based on the criterion's `importance`
and its content:

**essential criterion: evaluate exactly one dimension**
- If the criterion requires a specific time point or time span (localization),
  evaluate **temporal** only; set semantic to `"none"`.
- If the criterion requires content, meaning, causality, etc. (semantic), evaluate
  **semantic** only; set temporal to `"none"`.

**supporting criterion: evaluate both dimensions when applicable**
- If a `time_range` is provided: evaluate both **semantic** and **temporal**.
- If no `time_range` is provided: evaluate **semantic** only; set temporal to
  `"none"`.

### semantic

- **correct** = the response fully captures what the criterion requires; the key
  information is complete and accurate.
- **partial** = the response touches on part of the content, but key information is
  incomplete or contains errors.
- **wrong** = the response does not address what the criterion requires, or the core
  content is incorrect.
- **none** = the criterion is a localization task; semantic is not evaluated.

### temporal

- **correct** = the timestamp given by the model falls within the valid range (+/-5
  seconds tolerance around the target), or the time span given by the model overlaps
  the reference `time_range`.
- **wrong** = the model produced a timestamp or span, but it is outside the valid
  range and does not overlap.
- **missing** = the model did not produce a timestamp (the model is not required to
  do so).
- **none** = the criterion is semantic; temporal is not evaluated.

The model may use different timestamp formats. Normalize before comparing:
- `MM:SS` (e.g. `12:30`) = 12 minutes 30 seconds.
- Raw seconds (e.g. `522.640s`) = 8 min 42 sec = `08:42`.
- Out-of-range minutes (e.g. `64:00`) = `01:04:00`.

Format differences alone do not affect the judgment.

## Judgment Principles

- Focus on whether the core semantics are correct; different wording that expresses
  the same meaning is acceptable.
- Evaluate **semantic** and **temporal** independently: semantic judges only whether
  the content itself is correct, regardless of whether a timestamp is cited.
  Timestamp citation is judged under temporal.
- Each criterion is judged on its own requirements only. Details mentioned in other
  criteria apply to those criteria alone and do not affect this one.
- A criterion is satisfied as long as the core content is present and correct.
- Features listed in a criterion are examples; citing the key feature is sufficient.
- Keep `reason` and `summary` concise: 1-2 sentences each.

## Input

You will receive:
1. The question (open-ended QA).
2. A list of evaluation criteria (each with `text`, `importance`, and optionally
   `time_range`).
3. The model's response.

## Output Format

```json
{
  "criteria": [
    {
      "index": 0,
      "semantic": "none",
      "temporal": "correct",
      "reason": "The model timestamp 14:02 is within +/-5s of 14:03."
    },
    {
      "index": 1,
      "semantic": "correct",
      "temporal": "none",
      "reason": "The model correctly identifies the motive as loyalty formed from early allegiance to Sun Yat-sen."
    },
    {
      "index": 2,
      "semantic": "partial",
      "temporal": "wrong",
      "reason": "Semantic: the model mentions the vocal features but incompletely. Temporal: the cited timestamp is outside 00:41-00:45."
    }
  ],
  "summary": "A one-sentence summary of the overall quality of the response."
}
```
\end{promptbox}

\subsection{Models and APIs}
\label{app:models}

\begin{wraptable}{r}{0.5\textwidth}
\vspace{-\baselineskip}
\footnotesize
\caption{Construction-pipeline models.}
\label{tab:models}
\fittable{%
\begin{tabular}{@{}ll@{}}
\toprule
\textbf{Role} & \textbf{Model} \\
\midrule
Speech transcription (ASR) & \texttt{qwen3-asr-flash} \\
Acoustic captioning & \texttt{gemini-3.1-pro} \\
Language model (generation) & \texttt{qwen3.7-plus} \\
Audio verification & \texttt{qwen3-omni-flash} \\
Language model (judge) & \texttt{gpt-5.4} \\
\bottomrule
\end{tabular}
}
\end{wraptable}
Generation quality is enforced by the downstream verification and review stages (\S\ref{sec:construction-quality}), and the roles split by modality. The audio-facing stages (speech transcription, acoustic captioning, and audio verification) run on audio-capable models, while question generation and judging run on text-only language models. Table~\ref{tab:models} lists the models used in this release.

\section{Additional Results}
\label{app:sec-results}

\subsection{Performance Across Languages}
\label{app:language}

\textbf{The direction and magnitude of the language gap are model-dependent.} Splitting the leaderboard by language (Table~\ref{tab:language}), most models attain higher multiple-choice accuracy in English, though the magnitude varies substantially and the direction is not uniform. The gap is largest in open-ended generation for several open models that answer markedly better in English than in Chinese; Phi-4-Multimodal is the extreme case, scoring near zero on Chinese Rubric. The Chinese-origin Qwen-Omni family exhibits the opposite tendency: Qwen3.5-Omni Plus, the strongest model overall, is balanced on multiple-choice accuracy and slightly favours Chinese under Rubric. The gap can also be format-dependent within a single model: one Qwen3-Omni variant favours English strongly on multiple choice yet remains balanced under Rubric. Chain gaps are the smallest and directionless: sustaining the full P$\to$U$\to$R chain is hard in either language, so scores stay low and neither pulls far ahead.

\begin{table}[t]
\renewcommand{\yes}{\ding{51}}\renewcommand{\no}{\ding{55}}
\caption{Per-language scores (English vs.\ Chinese), each averaged over the three duration tiers. Formatting as in Table~\ref{tab:main-results}.}
\label{tab:language}
\centering
\setlength{\tabcolsep}{5pt}
\renewcommand{\arraystretch}{1.2}
\fittable{%
\footnotesize
\begin{tabular}{l @{\hspace{6pt}} c @{\hspace{5pt}} c *{2}{wc{0.6cm}} >{\cellcolor{lheaderbg}}wc{1.0cm} *{2}{wc{0.6cm}} >{\cellcolor{lheaderbg}}wc{1.0cm} *{2}{wc{0.6cm}} >{\cellcolor{lheaderbg}}wc{1.0cm}}
\toprule
& & & \multicolumn{3}{c}{\textbf{Accuracy (\%)}}
& \multicolumn{3}{c}{\textbf{Rubric (\%)}}
& \multicolumn{3}{c}{\textbf{Chain (\%)}} \\
\cmidrule(lr){4-6} \cmidrule(lr){7-9} \cmidrule(lr){10-12}
\textbf{Model} & \textbf{Size} & \textbf{Think} & \textbf{EN} & \textbf{ZH} & $\Delta_{\text{EN}-\text{ZH}}$ & \textbf{EN} & \textbf{ZH} & $\Delta_{\text{EN}-\text{ZH}}$ & \textbf{EN} & \textbf{ZH} & $\Delta_{\text{EN}-\text{ZH}}$ \\
\midrule
\multicolumn{12}{c}{\textbf{\textit{Open-source Models}}} \\
\midrule
\rowstripe Phi-4-Multimodal & 5.6B & \no & 38.3 & 30.0 & $+8.4$ & 23.3 & 4.9 & $+18.4$ & 10.0 & 9.2 & $+0.8$ \\
 &  & \no & 47.8 & 36.9 & $+10.9$ & 19.3 & 8.7 & $+10.6$ & 5.9 & 5.4 & $+0.6$ \\
\multirow{-2}{*}{AudioFlamingoNext} & \multirow{-2}{*}{8B} & \yes & 39.8 & 38.3 & $+1.5$ & 25.5 & 16.8 & $+8.8$ & 10.6 & 10.6 & $0.0$ \\
\rowstripe Voxtral-Mini & 3B & \no & 41.7 & 37.2 & $+4.4$ & 26.4 & 14.3 & $+12.2$ & 13.5 & 8.7 & $+4.8$ \\
Voxtral-Small & 24B & \no & 40.2 & 36.1 & $+4.1$ & 29.5 & 19.7 & $+9.8$ & 6.5 & 8.2 & $-1.7$ \\
\rowstripe  &  & \no & 43.1 & 39.3 & $+3.9$ & 20.1 & 14.6 & $+5.5$ & 8.3 & 10.9 & $-2.6$ \\
\rowstripe \multirow{-2}{*}{MOSS-Audio} & \multirow{-2}{*}{8B} & \yes & 49.4 & 46.3 & $+3.2$ & 19.5 & 18.6 & $+0.9$ & 14.3 & 12.8 & $+1.5$ \\
Baichuan-Omni-1.5 & 11B & \no & 45.9 & 40.9 & $+5.0$ & 15.5 & 12.0 & $+3.5$ & 12.4 & 15.0 & $-2.6$ \\
\rowstripe  &  & \no\textsuperscript{*} & 50.5 & 52.5 & $-2.0$ & 22.4 & 28.4 & $-6.0$ & 12.9 & 22.4 & $-9.5$ \\
\rowstripe \multirow{-2}{*}{Qwen3-Omni} & \multirow{-2}{*}{\makecell{30B\\(A3B)}} & \yes & 48.1 & 31.1 & $+17.0$ & 30.2 & 30.0 & $+0.2$ & 9.8 & 5.9 & $+3.9$ \\
\midrule
\multicolumn{12}{c}{\textbf{\textit{Closed-source Models}}} \\
\midrule
Gemini-2.5-Pro & -- & \yes & 63.3 & 60.9 & $+2.4$ & 43.2 & 39.4 & $+3.9$ & 22.0 & \second{25.6} & $-3.5$ \\
\rowstripe Gemini-3-Flash & -- & \no & 69.4 & 63.1 & $+6.3$ & \best{55.2} & \best{53.9} & $+1.3$ & 18.3 & 14.4 & $+3.9$ \\
Gemini-3.1-Pro & -- & \yes & \best{75.9} & \second{68.7} & $+7.2$ & \second{49.7} & 49.1 & $+0.5$ & \second{22.8} & 19.1 & $+3.7$ \\
\rowstripe Qwen3.5-Omni-Flash & -- & \no & 70.4 & 65.7 & $+4.6$ & 43.2 & 41.5 & $+1.7$ & 19.4 & 20.9 & $-1.5$ \\
Qwen3.5-Omni-Plus & -- & \no & \second{74.4} & \best{74.8} & $-0.4$ & 49.3 & \second{52.4} & $-3.0$ & \best{29.8} & \best{28.1} & $+1.7$ \\
\bottomrule
\end{tabular}%
}
\end{table}

\subsection{Performance Across Dimensions}
\label{app:tasktype}

\begin{figure}[t]
\centering
\includegraphics[width=\textwidth]{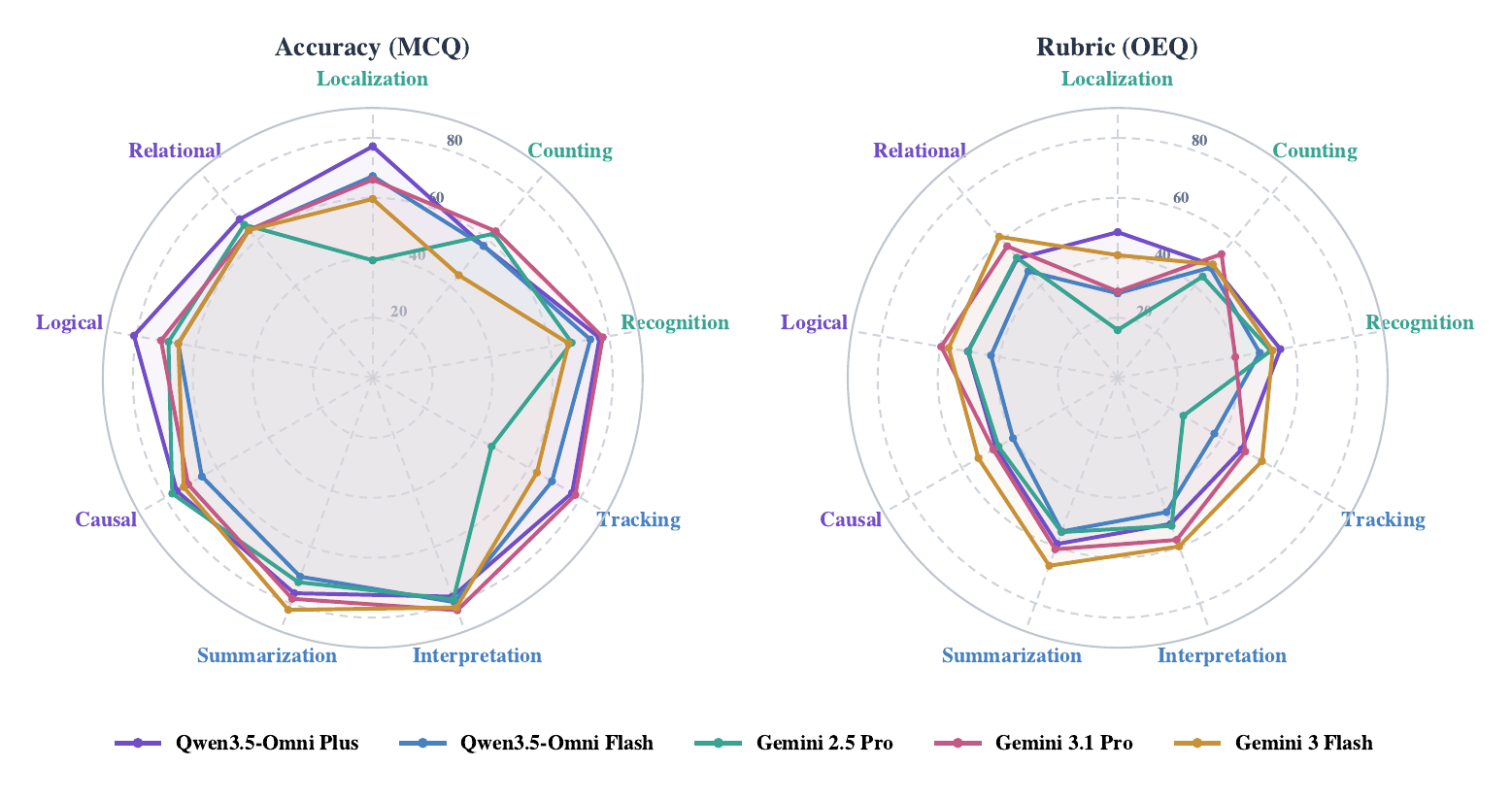}
\caption{Accuracy (left) and Rubric (right) across the nine dimensions for the five closed-source models. Axis labels are coloured by cognitive level: \textcolor{lteal}{\textbf{Perception}}, \textcolor{lblue}{\textbf{Understanding}}, \textcolor{lpurple}{\textbf{Reasoning}}. These three label colours coincide with three of the model line colours, but the labels sit on the rim and the lines in the field, so the two do not compete.}
\label{fig:dimension-radar}
\end{figure}

\textbf{Perception is the weakest level under both metrics, with localization and counting its low dimensions.} Averaged over the five models, perception scores below understanding and reasoning on both panels (Accuracy $63.8\%$, Rubric $43.6\%$), the same ordering as \S\ref{sec:exp-native}. Within it, localization and counting are the two low dimensions and recognition keeps pace with the stronger levels.

\textbf{Localization is the dimension multiple-choice most overstates, and counting the one it cannot lift.} The two lowest dimensions under Rubric, localization and tracking, both turn on placing events in time. Localization keeps only $53\%$ of its Accuracy score under open-ended grading, by far the steepest drop of any dimension, because multiple-choice lets a model pick a timestamp it cannot itself produce. Counting is the opposite: already low under Accuracy and barely lower under Rubric, a genuine capacity limit the easier format cannot lift. App.~\ref{app:content-timing} traces localization's collapse to the rubric's temporal axis. The same signature separates models: Gemini~2.5 Pro dips at localization and tracking while matching the field elsewhere, and Gemini~3.1 Pro closes that notch.

\subsection{Question Formats: Multiple-Choice vs.\ Open-Ended}
\label{app:mcq-oeq}

\textbf{On item-matched native questions, multiple-choice success is a weak proxy for grounded comprehension.} Because every native item is posed in both formats, we can measure this directly. For the five closed-source models we form the matched set $M$ of items scored in both formats and, for each item $i$, record its MCQ correctness $a_i \in \{0,1\}$ (Eq.~\ref{eq:accuracy}) and its open-ended Rubric score $r_i$ (Eq.~\ref{eq:rubric}). Splitting $M$ by the multiple-choice outcome into $M^{+} = \{i : a_i = 1\}$ and $M^{-} = \{i : a_i = 0\}$, we report the mean Rubric within each group,
\begin{equation}
\label{eq:rubric-pm}
\mathrm{Rubric}^{\pm} = \frac{1}{|M^{\pm}|} \sum_{i \in M^{\pm}} r_i .
\end{equation}
The overall Rubric is the weighted average of the two: with $p = |M^{+}|/|M|$ the MCQ-correct rate, $\mathrm{Rubric} = p\,\mathrm{Rubric}^{+} + (1-p)\,\mathrm{Rubric}^{-}$. We then summarize how far the multiple-choice outcome separates open-ended quality by the probability that a correct item outscores a wrong one on the rubric, i.e.\ the area under the ROC curve,
\begin{equation}
\label{eq:mcq-auc}
\mathrm{AUC} = \frac{1}{|M^{+}|\,|M^{-}|} \sum_{i \in M^{+}} \sum_{j \in M^{-}} \Big( \mathbf{1}[r_i > r_j] + \tfrac{1}{2}\,\mathbf{1}[r_i = r_j] \Big),
\end{equation}
where $0.5$ means the multiple-choice outcome is uninformative about the rubric score and $1$ that every correct item outscores every wrong one.

Pooled over the five closed-source models, open-ended answers to the very items answered correctly in multiple choice average only $\mathrm{Rubric}^{+} = 53.8$, far short of full credit, while the items answered incorrectly still average $\mathrm{Rubric}^{-} = 34.1$; the multiple-choice outcome separates the two by an AUC of just $0.65$, barely above the $0.5$ floor, and stays within $0.62$--$0.66$ across cognitive levels. Every closed-source model shows the pattern (Table~\ref{tab:mcq-shortcut}), though how it fails varies: Gemini~2.5 Pro separates its correct and wrong items most cleanly (AUC $0.71$) yet still reaches only $\mathrm{Rubric}^{+} = 51.1$ on the items it answers correctly, while Gemini~3 Flash pairs the highest $\mathrm{Rubric}^{+}$ ($59.9$) with the highest $\mathrm{Rubric}^{-}$ ($43.7$), making its multiple-choice outcome the least informative about open-ended quality (AUC $0.62$). No model both produces strong open-ended answers on the items it gets right and cleanly separates them from the ones it gets wrong.

\textbf{The clear cases divide into four cells, and disagreement runs mostly one way.} Scoring each open-ended answer as passing (Rubric $\geq 0.8$) or failing (Rubric $\leq 0.2$) and setting the partial middle aside (about $39\%$ of pairs), the matched (item, model) pairs split into four cells: both pass ($21\%$), both fail ($15\%$), correct only in multiple choice ($19\%$), and correct only open-ended ($6\%$). The multiple-choice-only cell outnumbers its mirror by roughly three to one. Figures~\ref{fig:mcq-case-both-right} through~\ref{fig:mcq-case-reverse} give one clean item from three of the cells, each evaluated on the same five models.

\textbf{The disagreement is a gap between recognizing an answer and producing it.} A perception localization item makes this concrete (Figure~\ref{fig:mcq-case}). With four candidate timestamps on offer, all five models select the right one; with the options withheld, none produces the true time and every open-ended answer scores zero on the localization rubric. Two models even quote the exact sentence that opens the queried event, locating the right moment in content while failing to place it in time.

\textbf{The other three cells show the effect is specific, and can even reverse.} When a model genuinely understands an item, the formats agree: on the acoustic-environment item of Figure~\ref{fig:mcq-case-both-right} all five pick the correct option and all five give the same correct open-ended description, scoring full credit. The reverse cell runs the other way (Figure~\ref{fig:mcq-case-reverse}): on a summarization item all five name the right answer open-ended, yet four of them choose a more dramatic distractor over that same answer in multiple choice, so the options mislead a model that can produce the answer. The both-failing cell holds items too hard for either format, such as an event no model finds in either form, where the shortcut simply does not arise.

\begin{table}[t]
\centering
\footnotesize
\setlength{\tabcolsep}{6pt}
\renewcommand{\arraystretch}{1.1}
\caption{Per-model average scores on the item-matched native set for the five closed-source models (matched items carry at least one essential criterion). $\mathrm{Rubric}^{\pm}$ and AUC are defined in Eq.~\ref{eq:rubric-pm} and Eq.~\ref{eq:mcq-auc}. Accuracy, Rubric, and $\mathrm{Rubric}^{\pm}$ are in percent; AUC is a probability in $[0,1]$.}
\label{tab:mcq-shortcut}
\begin{tabular}{lccccc}
\toprule
\textbf{Model} & \textbf{Accuracy} & \textbf{Rubric} & \textbf{Rubric$^{+}$} & \textbf{Rubric$^{-}$} & \textbf{AUC} \\
\midrule
\rowstripe Gemini 2.5 Pro & 62.0 & 41.2 & 51.1 & 25.1 & 0.71 \\
Gemini 3 Flash & 66.2 & 54.4 & 59.9 & 43.7 & 0.62 \\
\rowstripe Gemini 3.1 Pro & 72.4 & 49.3 & 54.9 & 34.4 & 0.65 \\
Qwen3.5-Omni Flash & 68.1 & 42.3 & 46.7 & 32.9 & 0.61 \\
\rowstripe Qwen3.5-Omni Plus & 74.5 & 50.8 & 56.0 & 35.8 & 0.66 \\
\midrule
Pooled & 68.6 & 47.6 & 53.8 & 34.1 & 0.65 \\
\bottomrule
\end{tabular}
\end{table}

\begin{figure}[t]
\begin{casebox}[gray]{Both correct: item L\_EN\_057, perception recognition}
\begin{alltt}
\cbhd{AUDIO}     1:05:51 total; a whole-recording judgment, no timestamp needed
\cbhd{QUESTION}  What best describes the acoustic environment of the recording?
\cbhd{OPTIONS}   A. outdoor, wind and traffic     B. dry close-mic studio (key)
          C. live stadium, distant crowd   D. reverberant hall with echo
\cbhd{CRITERIA}  essential: identifies a dry, clean, close-mic studio setting
          supporting: notes the absence of reverb, echo, and background noise

\cbhd{MULTIPLE-CHOICE}   scored against the key: all five models pick B (correct)
\cbhd{OPEN-ENDED}   options withheld, scored against the rubric: all five score full credit
  \textcolor{lblue}{Gemini 2.5 Pro}     dry studio, a podcast or treated room
  \textcolor{lblue}{Gemini 3 Flash}     a professional, dry studio setting
  \textcolor{lblue}{Gemini 3.1 Pro}     a small room or studio, clear voices, no echo
  \textcolor{lblue}{Qwen3.5-Omni Flash} quiet indoor, minimal ambience, no reverb
  \textcolor{lblue}{Qwen3.5-Omni Plus}  quiet indoor space, minimal reverberation
  all five describe the dry studio unaided: recognition here needs no options
\end{alltt}
\end{casebox}
\caption{A perception recognition item where the two formats agree: all five models pick the correct option, and all five give the same open-ended description for full credit. Model descriptions are abridged.}
\label{fig:mcq-case-both-right}
\end{figure}

\begin{figure}[t]
\begin{casebox}[gray]{Multiple-choice only: item S\_EN\_038, perception localization}
\begin{alltt}
\cbhd{AUDIO}     10:27 total; the 08:12 answer sits late in the recording
\cbhd{QUESTION}  At what time does the first customer testimonial begin speaking
          after the main presenter?
\cbhd{OPTIONS}   A. 08:24      B. 09:10      C. 08:12 (key)      D. 06:42
\cbhd{CRITERIA}  essential: names the switch to a distinct testimonial speaker at 08:12
          supporting: locates the presenter's speech just before the switch

\cbhd{MULTIPLE-CHOICE}   scored against the key: all five models pick C (correct)
\cbhd{OPEN-ENDED}   options withheld, scored against the rubric: all five score 0
  \textcolor{lblue}{Gemini 2.5 Pro}     04:55  "'We presently have four of these units ...'"
  \textcolor{lblue}{Gemini 3 Flash}     08:44  (quotes the same opening line)
  \textcolor{lblue}{Gemini 3.1 Pro}     01:14
  \textcolor{lblue}{Qwen3.5-Omni Flash} 08:31
  \textcolor{lblue}{Qwen3.5-Omni Plus}  05:28
  none states 08:12: each locates the event but cannot place it in time
\end{alltt}
\end{casebox}
\caption{A perception localization item where the two formats disagree: all five models pick the correct timestamp, none produces it open-ended, and two quote the right utterance at the wrong time. Model text is quoted verbatim and abridged.}
\label{fig:mcq-case}
\end{figure}

\begin{figure}[t]
\begin{casebox}[gray]{Open-ended only: item L\_EN\_028, understanding summarization}
\begin{alltt}
\cbhd{AUDIO}     1:01:10 total; the narrator's verdict sits in the closing segment
\cbhd{QUESTION}  In the concluding segment, what does the narrator call the primary
          missed opportunity that could have altered Marty Graham's path?
\cbhd{OPTIONS}   A. no mental-health treatment while employed (key)
          B. courts rejected the personality defense earlier
          C. no break in the foster-care abuse cycle before homelessness
          D. community missed signs of drug addiction
\cbhd{CRITERIA}  essential: names the failure to provide mental-health treatment
          supporting: ties it to thin support for parents of ill children

\cbhd{MULTIPLE-CHOICE}   scored against the key: all five miss it; four pick C, one picks D
\cbhd{OPEN-ENDED}   options withheld, scored against the rubric: all five score near full
  \textcolor{lblue}{Gemini 2.5 Pro}     failure to provide accessible mental-health care
  \textcolor{lblue}{Gemini 3 Flash}     lack of resources and early intervention
  \textcolor{lblue}{Gemini 3.1 Pro}     lack of care and support for the parents
  \textcolor{lblue}{Qwen3.5-Omni Flash} failure to give early intervention and support
  \textcolor{lblue}{Qwen3.5-Omni Plus}  lack of resources for parents raising such children
  all five name the treatment gap unaided; the options steer four of them to C
\end{alltt}
\end{casebox}
\caption{A summarization item where the disagreement reverses: all five miss the correct option, four choosing a distractor, yet all five name the right answer open-ended. Model answers are abridged.}
\label{fig:mcq-case-reverse}
\end{figure}

\subsection{Rubric Axes: Semantic vs.\ Temporal}
\label{app:content-timing}

\begin{figure}[t]
\centering
\includegraphics[width=\textwidth]{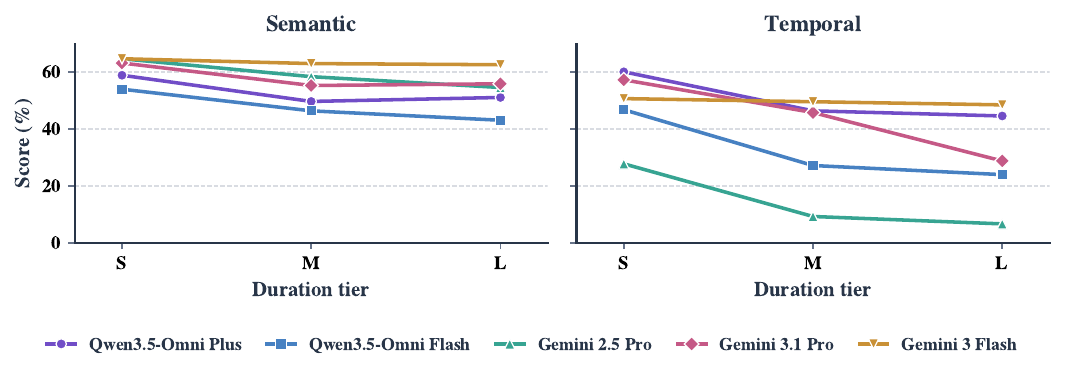}
\caption{Semantic vs.\ temporal accuracy by duration tier, per closed-source model, en+zh merged. \emph{Semantic\%} is the mean per-criterion content score; \emph{temporal\%} is timestamp accuracy.}
\label{fig:rubric-axes}
\end{figure}

\textbf{Open-ended performance is bounded by temporal grounding.} Each criterion carries a \emph{semantic} verdict (what happened) and a \emph{temporal} verdict (when), scored as in \S\ref{app:rubric}. Figure~\ref{fig:rubric-axes} plots both across the duration tiers, one line per model. Semantic accuracy is high and clustered ($43$--$65\%$) and drifts down only gently with length. Temporal accuracy is the weak, length-fragile axis: it starts lower and falls as the audio grows. The per-model shapes carry the story: Gemini~2.5 Pro is a pure localization failure, placing events in time correctly in single digits at $M/L$ while still describing them above $54\%$; Qwen3.5-Omni Plus stays the most grounded, its temporal tracking its semantic; Gemini~3.1 Pro holds temporal through $M$ then collapses at $L$; and Gemini~3 Flash alone keeps temporal flat across tiers, which is why its overall rubric score does not fall. The deficit is confident misplacement, not silence: pooled over the other four closed-source models, correct timing drops from $42\%$ at $S$ to $23\%$ at $L$ while wrong timing rises from $46\%$ to $67\%$, and outright omissions stay rare and flat ($10$--$13\%$).

The same gap explains the importance ordering. Every criterion is tagged \emph{essential} (the must-have answer, weight $1$) or \emph{supporting} (a corroborating detail, weight $0.5$), and models score higher on essential criteria ($50.7$, same pool) than supporting ones ($42.5$). That reads like models getting the core right and missing the extras, but it is not: semantic accuracy is nearly identical on the two levels ($58.8$ vs.\ $52.9$). What differs is timing load. Almost every supporting criterion demands a timestamp ($92\%$), whereas most essential criteria do not (only $27\%$ are graded on time), so the timing-heavy level inherits the weak temporal axis and scores lower. It is also where Gemini~3 Flash separates: its supporting score holds flat with duration, so its essential-supporting gap stays near zero.

\begin{table}[!t]
\centering
\footnotesize
\setlength{\tabcolsep}{4pt}
\renewcommand{\arraystretch}{1.0}
\caption{Per-config Anchor-QA link rates and the grounding split, pooled over tiers and languages. P/U/R are per-link accuracy; Chain the truncated score (Eq.~\ref{eq:chain}). $U\mid$loc.\ and $U\mid$miss.\ are understanding accuracy with the anchor located or missed (read as located $-$ missed).}
\label{tab:anchor-config}
\begin{tabular}{lcccccc}
\toprule
\textbf{Config} & \textbf{P} & \textbf{U} & \textbf{R} & \textbf{Chain} & \textbf{$U\mid$loc.} & \textbf{$U\mid$miss.} \\
\midrule
\rowstripe Gemini 2.5 Pro & 38.3 & 48.6 & 42.5 & 23.8 & 54.3 & 45.0 \\
Gemini 3 Flash & 29.7 & 46.9 & 36.1 & 16.4 & 45.8 & 47.4 \\
\rowstripe Gemini 3.1 Pro & 38.3 & 48.1 & 39.2 & 20.9 & 42.8 & 51.4 \\
Qwen3.5-Omni Flash & 38.1 & 39.7 & 35.8 & 20.2 & 41.6 & 38.6 \\
\rowstripe Qwen3.5-Omni Plus & 47.8 & 47.2 & 41.7 & 29.0 & 54.1 & 41.0 \\
\bottomrule
\end{tabular}
\end{table}

\begin{figure}[t]
\centering
\includegraphics[width=0.85\textwidth]{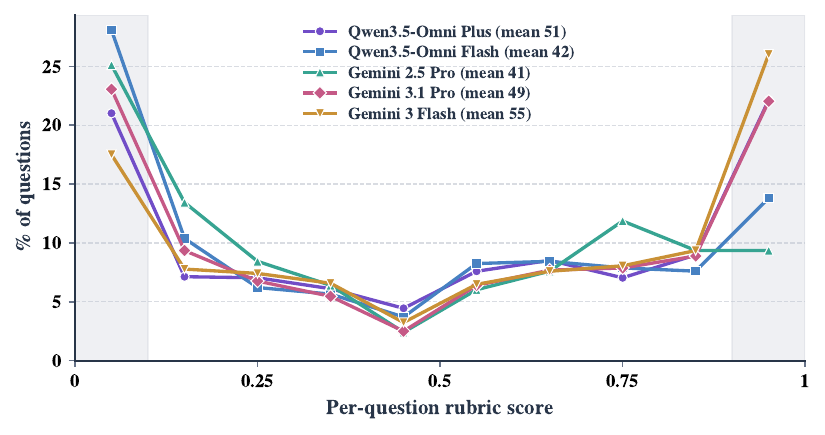}
\caption{Per-question rubric score distributions, pooled over the S/M/L tiers, per closed-source model (en+zh merged): the \% of questions in ten equal-width bins over $[0,1]$, pole bins $[0,0.1)$ and $[0.9,1.0]$ shaded. The legend gives each model's mean.}
\label{fig:rubric-dist}
\end{figure}

Underneath these means, the per-question rubric scores are close to all-or-nothing (Figure~\ref{fig:rubric-dist}). Every model's distribution is U-shaped: the two largest bins are the extremes, $[0,0.1)$ and $[0.9,1.0]$, and the middle bins are the emptiest. Models differ mainly in how the mass splits between the poles. The strongest place the most at the top: Gemini~3 Flash scores $26\%$ of questions in the top bin against $18\%$ in the bottom, whereas the confident misplacer Gemini~2.5 Pro reaches the top bin on only $9\%$ and piles $25\%$ at the bottom.

\subsection{Chain Scoring: Grounding vs.\ Guessing}
\label{app:anchor-analysis}

\textbf{Understanding and reasoning are answered without locating the anchor.} On each chain we split understanding and reasoning accuracy by whether the perception link was correct, that is, whether the anchor was located or missed on that same chain; when the missed-anchor rate matches the located rate, the credit does not come from grounding. Pooled over the five closed models (Table~\ref{tab:anchor-config}), understanding is answered $44.9\%$ of the time on chains whose anchor was never located, against $48.1\%$ when it was, a three-point grounding premium; reasoning is similarly flat ($37.5$ vs.\ $41.6$). The split does separate models: understanding tracks grounding most for Qwen3.5-Omni Plus ($54.1$ located vs.\ $41.0$ missed, $+13.1$) and inverts for Gemini~3.1~Pro thinking ($-8.6$), which answers understanding \emph{more} often when it missed the anchor, the signature of a guesser.

\textbf{First-error truncation floors a guesser and credits only grounding.} The text-only solvers (\S\ref{sec:exp-setup}) trace what the chain score responds to. A blind solver clears each link $22$--$34\%$ of the time on its own, yet scores only $10.4$ on the chain, because ungrounded guesses cannot be collected into a correct prefix. Adding the ASR transcript barely moves it ($12.6$). Only injecting the anchors into the caption (the oracle of Figure~\ref{fig:anchor-chain}) recovers the chain, to $61.4$. The score therefore tracks temporal grounding specifically, not verbal content, and resists the per-link guessing that inflates the link rates.

\begin{wraptable}{r}{0.45\textwidth}
\centering
\footnotesize
\setlength{\tabcolsep}{4pt}
\renewcommand{\arraystretch}{1.0}
\caption{Anchor-QA link rates and chain score by anchor operation, pooled over the five closed models, tiers, and languages. Addition inserts a sound event, Deletion silences a span, Modification alters acoustic attributes. P/U/R are per-link accuracy; Chain the truncated score (Eq.~\ref{eq:chain}).}
\label{tab:anchor-operations}
\begin{tabular*}{\linewidth}{@{\extracolsep{\fill}}lcccc}
\toprule
\textbf{Operation} & \textbf{P} & \textbf{U} & \textbf{R} & \textbf{Chain} \\
\midrule
Addition & 38.7 & 49.3 & 34.0 & 22.6 \\
Deletion & 32.8 & 45.0 & 42.5 & 19.2 \\
Modification & 43.8 & 44.0 & 40.7 & 24.3 \\
\bottomrule
\end{tabular*}
\end{wraptable}
\textbf{The guessing signature is strongest for the operation that is hardest to locate.} The three anchor operations differ in how findable they leave the target (Table~\ref{tab:anchor-operations}). Pooled over the five closed models, perception is highest for Modification ($43.8$), then Addition ($38.7$), then Deletion ($32.8$): a silenced span is the hardest anchor to find and an altered but still-present one the easiest, an ordering that holds for every model and every duration tier. Yet reasoning peaks under Deletion ($42.5$) despite its worst perception, the one operation where a higher layer scores above its own perception link, and its chain score stays lowest ($19.2$): the operation hardest to ground is exactly where the higher layers answer without it.

\section{Limitations}
\label{app:sec-limitations}

The two question paths are complementary, yet each carries its own constraint. For Native QA, coverage is limited by caption fidelity: items can only draw on what the caption records, and the bound tightens on the longest audio, where omissions are hardest to catch and human review cannot recover what was never written. For Anchor QA, synthetic anchors provide deterministic ground truth, but confine questions to the grounding-to-reasoning chain template, sacrificing the free-form diversity that Native QA offers. Beyond these design tradeoffs, the benchmark covers two languages (English and Chinese) and in-the-wild speech; low-resource languages, cross-lingual questions, and long music or environmental recordings lie outside its scope.

\end{document}